\documentstyle[preprint,aps,12pt]{revtex}
\begin{document}
\tighten
\bibliographystyle{prsty} 
\input epsf.sty 

\title{ Potential of a neutral impurity in a large $^4$He cluster}
\author{ Kevin K.~Lehmann\footnote{1998 Visiting Fellow.  Permanent address: 
Department of Chemistry,
Princeton University, Princeton NJ 08544}}
\address{JILA,\\ University of Colorado and National Institute of Standards
and Technology,\\ Boulder, CO 80309--0440}

\date {To be published in \em Molecular Physics}

\maketitle 

\begin{abstract} 
This paper presents an analysis of the motion of an neutral impurity species
in a nanometer scale $^4$He cluster, extending a previous study of the
dynamics of an ionic impurity.  
It is shown that for realistic neutral impurity--He potentials, such
as those of SF$_6$ and OCS, the impurity is kept well away of the
the surface of the cluster by long range induction and dispersion
interactions with He, but that a large number of `particle in a box'
center of mass states are thermally populated. It is explicitly 
demonstrated how to calculate the spectrum that arises from the
coupling of the impurity rotation and the center of mass motion, and
it is found that this is a potentially significant source of
inhomogeneous broadening in vibration--rotation spectra of anisotropic 
impurities.  Another source of inhomogeneous broadening is the 
hydrodynamic coupling of the rotation of the impurity with the 
center of mass velocity.  A quantum hamiltonian to describe this effect
is derived from the classical hydrodynamic kinetic energy of an
ellipsoid.  Simple analytic expressions are derived for the resulting
spectral line shape for an impurity in bulk He, and the
relevant matrix elements derived to allow fully quantum calculations of
the coupling of the center of mass motion and rotation for an impurity
confined in a spherical He cluster.
Lastly, the hydrodynamic contribution to the impurity effective moment of inertia
is evaluated and found to produce only a minor fractional increase. 

\end{abstract}

\section{Introduction}\label{sec-Intro}

The last few years have seen dramatic advances in the spectroscopy of
atoms and molecules attached to large He clusters~\cite{Whaley98}. 
These clusters provide a unique environment for a spectroscopy
which combines many of the attractive features of both
high resolution gas phase spectroscopy and traditional
matrix spectroscopy~\cite{Lehmann98}.  
These include the ability to obtain rotationally resolved spectra of
even very large molecules such as SF$_6$~\cite{Hartmann96}
and (CH$_3$)$_3$SiCCH~\cite{Callegari_up} (though with effective rotational
constants only 1/3 -- 1/5 as large as those of the gas phase molecule), 
and the ability to form and stabilize
extremely fragile species~\cite{Higgins96a}, 
including high spin states of molecules.  
Particularly interesting is the recent demonstration by Grebenev, 
Toennies, and Vilesov~\cite{Grebenev98}
 that the free rotation is a direct consequence of the
Boson character of $^4$He, and can be viewed as a microscopic
version of the famous Andronikashvili experiment~\cite{Andronikashvili46}
that was used to measure the
superfluid fraction in bulk liquid He.  

Despite rapid progress, many fundamental questions remain about
spectroscopy in this environment.  One such important question is
the origin of the linewidth observed in ro--vibrational spectra.  
These linewidths vary with species, from 150 MHz reported for the 
R(0) line of the OCS
$\nu_3$ fundamental band~\cite{Hartmann_thesis}, 
to 1.5 cm$^{-1}$ reported for H$_2$O~\cite{Frochtenicht96}.  
Very recently, pure microwave spectra have been observed for
impurity molecules (CH$_3$CN and HCCCN) which have linewidths on the
order of $1$\, GHz~\cite{Reinhard98}.  
Since this is comparable to linewidths observed in
ro--vibrational transitions, it is unlikely that vibrational dephasing, 
which is the dominant line broadening mechanism in the spectra of impurities
in classical liquids, also plays a dominant role in He cluster spectroscopy.
Likewise, vibrational population relaxation, which has been invoked
as the source of line broadening, obviously cannot contribute to the
broadening in the case of microwave excitation.

Given the fact that the He clusters are uniquely fluid, even down to zero
temperature, it is natural to suppose that the impurity spectra will not 
display inhomogeneous effects.  Variations in local binding sites tend to 
dominate the linewidth of spectra in low temperature crystal or glass
hosts.  In a liquid environment, local changes in solvation lead to
dynamic fluctuations in the spectral intervals, and thus can lead to
dephasing but not static inhomogeneous effects.  Since the time scale
for solvation fluctuations, due to the large zero point kinetic energy 
of He atoms in the bulk, is expected to be much faster than the 
dephasing times observed in most ro--vibrational spectra, one can expect
the effects of the fluctuating solvation to be strongly motionally averaged.
More fundamentally, it is far from clear that we expect pure vibrational
dephasing at all given the low temperature of He clusters.  Taking the
model of a crystalline solid, one expects dephasing to arise from interactions
with thermal excitations.  Since, as will be discussed below, the only
excitations thermally excited are surface waves, and these are expected to
interact extremely weakly with impurities, it is difficult to see how
dephasing can be important.

The fact that the existing experiments produce a broad distribution
of cluster sizes suggests an inhomogeneous contribution to the line broadening 
from a size dependent shift in the absorption.  However, it is likely that 
this effect
is also not dominant in most spectra.  
First, the over all shift of ro--vibrational
line from the gas phase 
to He clusters is fairly small, typically  at most
a few cm$^{-1}$~\cite{Whaley98},
so the spread
in shifts as one approaches the bulk limit is expected to be small.
Below, an estimate of the expected shift of the SF$_6$ 
$\nu_3$ vibrational transition due to the finite size of the He cluster
will be given.
Further, the
experimentalist has some control over the mean of the cluster size distribution.
In previous work on SF$_6$, it was found that the spectral shift and temperature
was essentially constant for He clusters of $>10^3$ 
He atoms~\cite{Hartmann_thesis}.  

An additional source of inhomogeneous environment needs to be considered, the
distribution of the position and orientation of the impurity relative to the
surface of the cluster.  That this could be relevant is suggested 
by the following simple calculation.  
Consider an impurity with an effective mass (including
hydrodynamic contributions) of 50 atomic mass units (u)
in equilibrium with a He cluster, which is known to be at 
$0.38$\,K~\cite{Hartmann96}.  This
molecule will have an root mean squared (RMS) velocity of $14\, \rm m\,s^{-1}$.
In a cluster of diameter $10$\,nm, which corresponds to $\approx 11,000$\, 
He atoms, it will take $\approx 0.7$\,ns for the impurity to cross the
cluster.  If the scattering of the impurity from the surface of the
cluster leads to re--orientational dephasing, the resulting line broadening
will be $\approx 500$\, MHz, comparable to what has been observed in 
ro--vibrational transitions~\cite{Whaley98}.  An additional source of 
dephasing is the
hydrodynamic coupling of the molecular orientation to the molecular velocity,
which will be described in more detail below.

The simplest treatment for the motion of the impurity molecule is as a 
`particle in a box'.  This was considered in an earlier paper by Toennies
and Vilesov~\cite{Toennies95}.
That treatment was unsatisfactory on several grounds however.  First, as
we will demonstrate below, the true potential is not really `flat' inside
the box relative to the low value of the thermal energy.  
Second, and perhaps more importantly, they did not attempt to calculate the
intensity of transitions in this model.  Since the effective potential is
almost the same in the upper and lower vibrational states, there are
small Frank--Condon factors for transitions that change the 
center of mass quantum numbers of the impurity, and as such, changes in
energy of the particle in a box levels are not directly observable in the
spectrum of a neutral impurity. 

This paper will present a model that allows
calculation of the effective potential for the
motion of an impurity that is
free to move about in a spherical liquid He droplet of radius R.  
A thorough analysis of the liquid drop model for pure He clusters
was provided by Brink and Stringari~\cite{Brink90}.  
The present calculation will be based upon long range interactions and treat
the He as a continuum incompressible liquid of constant number density, 
$\rho$, which
we will take as the known value of bulk liquid He.  
While such a model cannot be expected to correctly describe the interaction
of the impurity with the first few solvation shells, this solvation
structure is expected to be effectively constant as long as the impurity 
does not too closely approach the surface of the cluster.  For realistic
values of the impurity--He long range interaction, the energetics are
effective at keeping the impurity from this difficult region.
Further, since the impurity is kept away from the edge of the cluster, 
the fact that the He cluster has a diffuse ($\approx 7\, {\rm  \AA}$) interface with
vacuum~\cite{Lurio93} will
not be expected to significantly effect the impurity potential.
Further, we can, with good approximation, keep only the lowest order terms
in the expansion of the long range He--impurity interaction in inverse 
powers of distance.    

The paper is organized as follows. 
Section~\ref{sec-SF6}  will consider a neutral, isotropic impurity,
with SF$_6$ used as an example.  
Sections~\ref{sec-Linear} and \ref{sec-HCN} will
consider the case of a symmetric top or linear molecule impurity, with HCN used as
an example.  Section~\ref{sec-Linear} will derive the effective
potential and apply a classical model for calculation of the resulting
spectra.  Section~\ref{sec-HCN} will derive the matrix elements
needed for a fully quantum treatment of the interaction of the center of mass
motion and the rotations of the impurity for
a linear molecule, leading to a calculation of the 
expected spectral structure for the case of the HCN R(0) rotational line.
Section~\ref{sec-Ripplons} will consider the effect of the interaction of an 
impurity with the internal modes of the cluster, deriving the expected shifts 
produced by the interaction with the ripplons, which are quantized surface 
capillary waves.  These interactions are found to be completely negligible
contributions to the linewidth of transitions.  
Section~\ref{sec-Hydro} will consider Hydrodynamic effects and 
derive the effective hamiltonian and
matrix elements for the hydrodynamic coupling of the rotation and center of
mass momentum of the impurity.  When this model is applied
to the HCN R(0) line, poor agreement is found with experiment.  
However, it will be shown in Section~\ref{sec-OCS} that the combination of the
anisotropic potential terms and hydrodynamic coupling leads to a prediction
for the lineshape of the R(0) line in the OCS ro--vibrational spectrum that 
is in excellent agreement with that observed in Helium clusters and 
reported by Toennies Group~\cite{Grebenev99}.
Section~\ref{sec-Inertia} will examine the hydrodynamic contribution to the reduction
in rotational constant for an impurity, treated as an ellipsoid. It is
found that the hydrodynamic kinetic energy only makes a minor contribution
to the observed increase in moment of inertia of impurity molecules.
Lastly, section~\ref{sec-Summary} will provide an overview and summary of this work.

\section{Isotropic Neutral Impurity, such as SF$_6$, in a Helium Cluster}\label{sec-SF6}

In a previous paper~\cite{Lehmann99a}, the potential of an ionic
impurity in a helium cluster was derived from the ion--induced dipole
interaction of the impurity in Helium.
The present paper will consider the case of a neutral impurity. 
As will be shown, this potential is dominated by the leading power
in the long range interaction between the impurity and atomic helium.
In the present section, the case of an impurity
that has an isotropic long range interaction with He will
be considered.  This will
be the case for impurity molecules of tetrahedral or octahedral 
symmetry as well as for atomic impurities.  
In the next section, the case of an anisotropic long
range interaction, as expected for a linear molecule impurity, will
be treated.

As in our previous paper, we will take as a reference of energy not an
isolated impurity atom or molecule, but the impurity in bulk He with constant
He number density, $\rho$.  In this way, the difference between an impurity
inside the cluster and our reference state is the absence in the later case
of He outside the cluster.  We will also implicitly assume that the impurity is
far enough away from the surface of the He cluster that we treat
the interaction of the impurity with this `missing' He as pairwise
additive and 
we can use only the long range terms of the potential.

Let the interaction energy of the impurity with the He be a
van der Waals attraction at long range, with a long range
form of :
\begin{equation}
V(r) = - \frac{C_6}{r^6} - \frac{C_8}{r^8} \ldots \label{eq:V}
\end{equation}
Where $C_6$ and $C_8$ are the long range dispersion interaction
constants and r is the distance from the center of mass of the
impurity to the He atom.  
For SF$_6$, the lowest order anisotropic term
appears at the $r^{-10}$ level. 

Consider the impurity to be displaced
from the center of a He cluster of radius $R$ by
a distance $a$.  
The $z$ axis is defined as parallel to displacement vector of the
impurity from the center of the cluster, which provides the 
origin for the coordinate system. 
Let $r(\theta)$  be the distance from the impurity to the droplet surface
at a polar angle $\theta$ measured {\it from the impurity}. 
Basic trigonometry gives:
\begin{equation}
R^2 = (a + r \cos(\theta) )^2 + r^2 \sin(\theta)^2 = 
a^2 + r^2 + 2 ra \cos(\theta)
\end{equation}
from which we can derive:
\begin{equation}
r(\theta) = \sqrt{R^2 - a^2 \sin(\theta)^2} - a \cos(\theta) \label{eq:r(theta)}
\end{equation}
We can `sum' up the missing interaction for all the He atoms absent 
(compared to the bulk phase) due to the finite extent of the cluster,
using the interaction potential given above in Eq.~\ref{eq:V}:
\begin{eqnarray}
\Delta E (a,R) & = & \int_0^{\pi} \, \int_{r(\theta)}^{\infty}
\left[ \frac{\rho\, C_6}{r'^6} + \frac{\rho\, C_8}{r'^8} \right]
\, 2 \pi\, r'^2 \sin(\theta) \, dr' \, d\theta 
\nonumber \\
 & = & V_2^0 F_2(a/R) + V_3^0 F_3(a/R)  \label{eq:E_SF6} \\
V_2^0 & = & \left(\, \frac{4\, \pi\, \rho\, C_6}{3\, R^3}\, \right) \label{eq:V_2}\\
F_2(y) & = & \left[ 1 - y^2 \right]^{-3} 
\approx  1 + 3 y^2 + 6 y^4 + 10 y^6 + \ldots \label{eq:F_2}\\
V_3^0 & = & \left(\, \frac{4\, \pi\, \rho\, C_8}{5\, R^5} \right) \label{eq:V_3} \\
F_3(y) & = & \frac{ 1 + \frac{5}{3} y^2}{ (1 - y^2)^5}
\approx 1 + \frac{20}{3} y^2 + \frac{20}{3} y^4 + 60 y^6 \ldots \label{eq:F_3}
\end{eqnarray}
We will consistently use $y = a/R$ in this paper and refer to $y$ as
the normalized displacement of the impurity.  We start with
label $F_2$ to avoid confusion with the function $F_1$ defined in the 
paper on ionic impurities.

\begin{figure}
\centerline{\epsfbox{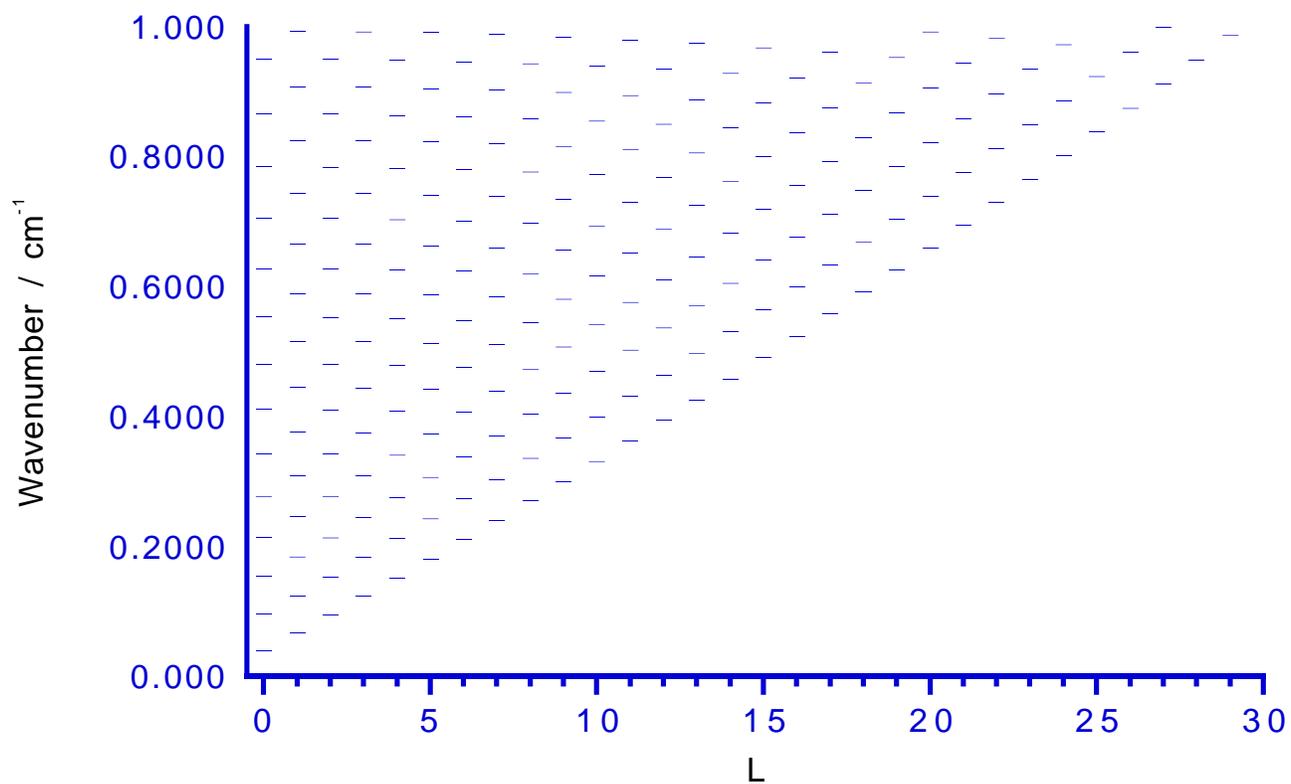}}
\vspace{1in}
\label{Energy_Levels}
\caption{Center of mass motion energy levels for SF$_6$ in a
He cluster of radius 3 nm.  All levels below 1 cm$^{-1}$ are plotted.
The ordinate is L, the total orbital angular
momentum quantum number with states of increasing n, the number of nodes
in the radial direction, increasing vertically in each column.  }
\end{figure}

\begin{figure}
\centerline{\epsfbox{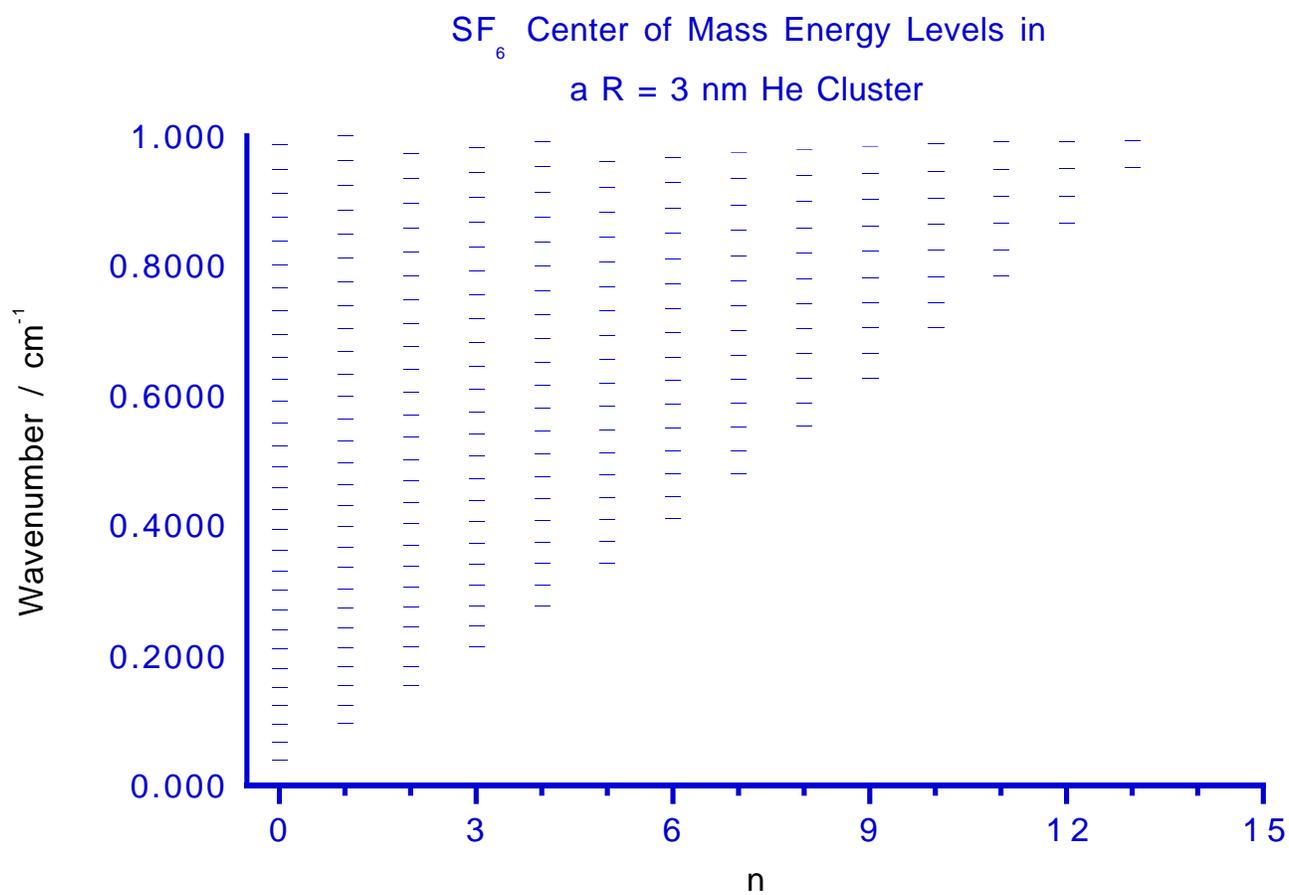}}
\vspace{1in}
\label{Energy_Levelsb}
\caption{Same as figure 1, but with n as the
ordinate and states of increasing L increasing vertically in each column.}
\end{figure}

\begin{figure}
\centerline{\epsfbox{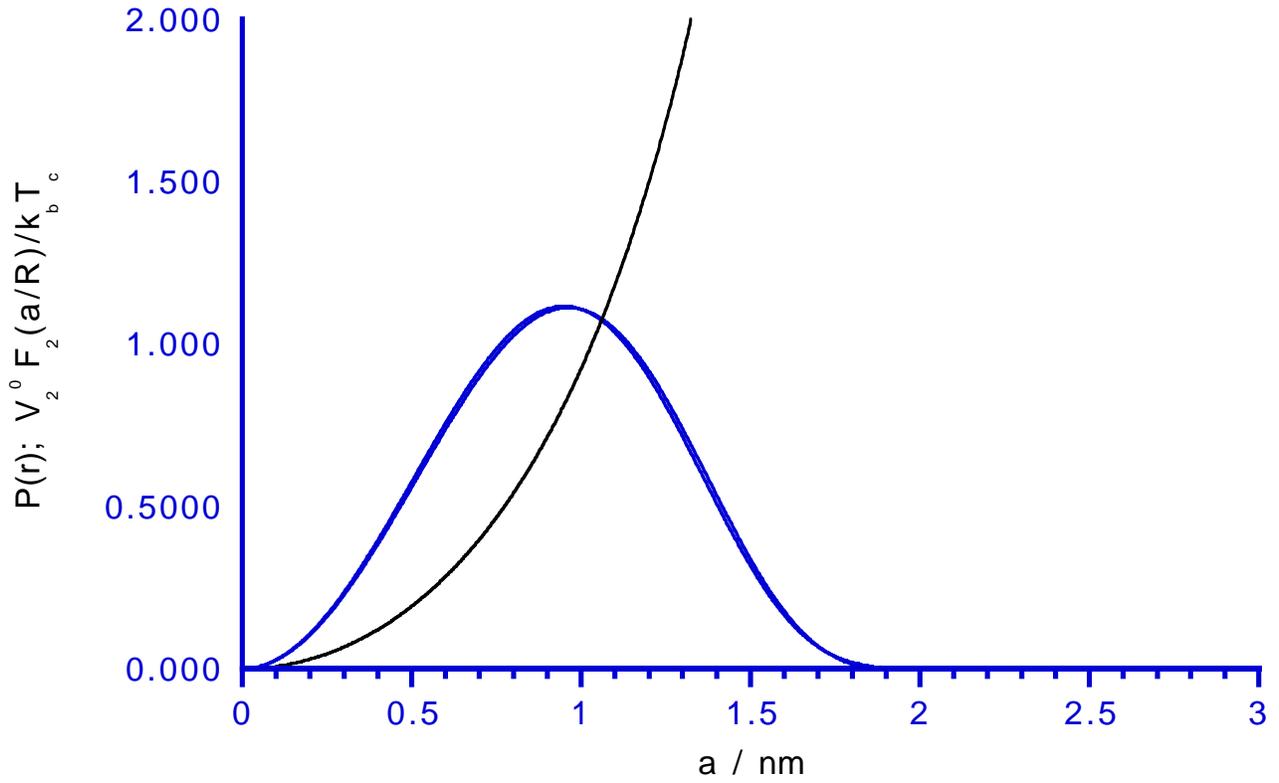}}
\vspace{1in}
\label{Radial_Density}
\caption{Comparison of the classical and quantum Radial Distribution Functions, P(a),
for SF$_6$ in
a 3 nm He cluster.  The Quantum distributions has been calculated from the
set of all eigenstates with wavenumber less than 2 cm$^{-1}$.  Also plotted,
is the isotropic potential, U(a), that confines the SF$_6$ in the cluster.
}
\end{figure}

In atomic units, the $C_6$ and $C_8$ coefficients are typically of
the same magnitude~\cite{Standard85}, and thus we expect $V_3^0$ to be
smaller than $V_2^0$ by a factor
on the order of a$_0^2/R^2 \approx 10^{-4}$
(a$_0$ is the Bohr radius), and thus we will neglect it.
Using the $C_6$ coefficient for SF$_6$--He equal to $35\, {\rm E_h} {\rm a}_0^6
= 3.35 \cdot 10^{-78}\, {\rm J\,m}^6$~\cite{Pack82} ($\rm E_h$ is
the Hartree, the atomic unit of energy),
$\rho = 0.022\, {\rm atoms \, \AA}^{-3}$ (the density of liquid He at
low pressure and temperature),
and $R = 3$\, nm (which corresponds to the size of a 
cluster containing $\approx 2500\,$He atoms), we calculate
$V_2^0 = 0.572\,{\rm cm}^{-1}\cdot h \, c \approx 2 \,  k_b \, T_{\rm c}$.
$k_b$ is Boltzmann's constant and $T_{\rm c} = 0.38$\,K is the temperature
of nanometer scale He clusters.  
If we assume that the effective mass for translational motion of the impurity
is $M_{\rm eff} = 186$\, u (the SF$_6$ plus $10$ He atoms
to include hydrodynamic effects and some `sticking' of
solvated He atoms), we get a harmonic frequency for vibration of the impurity
about the center of the He cluster
of $790$\, MHz.  
Figures~\ref{Energy_Levels} and \ref{Energy_Levelsb} show the
energy levels of this potential below $1\, {\rm cm}^{-1} \approx 4 \, k_b T_{\rm c}$, 
calculated using
the Numerov-Cooley method~\cite{Tellinghuisen88}.  
Inspection of the Energy as a function of n (radial) and L (angular momentum)
quantum numbers shows that the thermally well populated energy levels have
a spectrum much more like that of a particle in a three dimensional
harmonic well than a particle in a spherical `box'.
Figure~\ref{Radial_Density}
contains a plot of the
probability density for finding the impurity a distance $a$ from the
from the center of a $R = 3$\,nm cluster, calculated both from the 
set of eigenfunctions and from the classical Boltzmann
distribution.  We see an excellent agreement of the two,
confirming the classical character of this motion even at 
$T_{\rm c}$.  We see that the probability density is largely
localized in the region with $a$ less than about half the
cluster radius.  The RMS value for the displacement of the
impurity from the cluster center is $0.96$\,nm.  
The highly classical nature of the 
thermal motion can also be seen by comparing the thermodynamic 
functions calculated from the exact quantum energy levels and for 
classical motion with the same potential.  This comparison is shown
in table 1. Also included in this table is the mean squared orbital
angular momentum (calculated in units of $\hbar^2$), 
calculated as the thermally averaged value
of $L(L+1)$, where $L$ is the orbital angular momentum quantum number of the
impurity motion around the center of the cluster,
which is compared to the classical thermal mean value of
$< \mbox{\boldmath $L$}^2 > = 2 M_{\rm eff} k_b T_{\rm c} <a^2>/ \hbar^2$.  
The table also contains the
thermodynamic averages for a cluster of $R = 5$\,nm as well as the
corresponding properties of the ripplons (to be discussed below).  
It can be seen that the average energy and heat capacity are 
intermediate between those of a true particle in a box and a
particle in a Harmonic well.  The partition
function for translation grows faster than the volume of the cluster,
a factor of $15.3$ compared to $4.6$ when the radius of the cluster is increased
from $3$ to $5$\,nm.  This reflects the fact that in a larger cluster,
the impurity gets to thermally sample a larger fraction of the cluster volume,
as demonstrated by the RMS thermal displacement, $\sqrt{a^2}$.

\section{Symmetric top or Linear Molecule Impurity in the Helium Cluster}\label{sec-Linear}

We will now assume we have an impurity with an
anisotropic interaction with He.  If the impurity is in the center
of the cluster, it will experience an isotropic net potential, due
to the symmetry of the distribution of He atoms.  However, as the impurity
moves away from the center, it will experience an anisotropic potential.
This anisotropy will cause a coupling between the center of mass motion
of the impurity and its rotational motion, and can be expected to result
in orientational dephasing.  It is of interest to point out that the
thermal mean value of the orbital angular momentum 
({\boldmath$L$}) of the impurity in the
cluster $\approx 20\, \hbar$ is much greater than the thermally populated
value of rotational angular momentum of the impurity, ({\boldmath $j$}), and as a consequence 
the procession of {\boldmath $j$} about the total angular momentum 
{\boldmath $J$} would be
expected to completely dephase the average dipole moment, and lead to
line broadening.   

The long range interaction of a neutral, symmetric top impurity with He can be 
written
as the sum of two parts, one due to induction and the other dispersion.
The former is dominated  by the dipole induced dipole interaction for
an impurity with a permanent dipole moment.
We can write the potential in the following form~\cite{Atkins96}:
\begin{equation}
V(r,\phi) = - \frac{C_{60}}{r^6} - \frac{C_{62}}{r^6} P_2(\cos(\phi))
-\frac{C_{71}}{r^7} P_1(\cos(\phi)) - \frac{C_{73}}{r^7} P_3(\cos(\phi))
\label{eq:V(r)}
\end{equation}
where $P_n$ are the Legendre polynomials, and $\phi$ is the angle
between the vector from the center of mass of the impurity to the He, and
the symmetry axis of the impurity.
The induction contributions to these coefficients can be 
calculated~\cite{Atkins96}
from the dipole moment, $\mu$, and quadrupole moment, $\Theta$, of the
impurity and the polarizability, $\alpha$, of He:
\begin{equation}
C_{60}^{\rm ind}\, =\, C_{62}^{\rm ind} \, = \,
\frac{\alpha\, \mu^2}{(4\,\pi\,\epsilon_0\,)^2} \label{eq:Cind}
\end{equation}
\begin{equation}
C_{73}^{\rm ind} \, =\, \frac{2}{3}\, C_{71}^{\rm ind} \, = \,
\frac{12\, \alpha \, \mu \, \Theta}{5 \, (4\, \pi\, \epsilon_0\,)^2 }
\end{equation}
The dispersion contributions to the long range interactions can be
written as integrals over the imaginary frequency dependence of the 
polarizability
of He and the polarizability and dipole-quadrupole
polarizabilities of the impurity~\cite{Dalgarno67}.
Polarizablility is higher parallel than perpendicular to bonds. 
As a result, for linear molecules and most prolate symmetric tops,
$C_{62} > 0$, which means that for such impurities, the interaction is
lower in energy when the He-impurity displacement is parallel rather than
perpendicular to the impurity symmetry axis.

As an example, we will consider the case of HCN as the impurity.
Using the parameters $\alpha({\rm He}) = (4\,\pi\,\epsilon_0)\cdot 1.383\, a_0^3$,
$\mu({\rm HCN}) = 1.174\, e\, a_0 = 2.98\,$Debye, and $\Theta({\rm HCN})
= 1.777\, e\,a_0^2$~\cite{Atkins96}, we get that the 
induction contribution to the long
range coefficients are: 
$C_{60}^{\rm ind}\, =\, C_{62}^{\rm ind} \, = \, 1.14 \; {\rm eV \AA}^6$
and
$C_{73}^{\rm ind} \, =\, \frac{2}{3}\, C_{71}^{\rm ind} \, =
 \, 2.19 \;{\rm eV \AA}^7$.  
The dispersion contributions to the coefficients have been calculated
by Atkins and Hutson~\cite{Atkins96} to be
$C_{60}^{\rm disp} = 7.81 \;{\rm eV \AA}^6$,
$C_{62}^{\rm disp} = 1.11 \;{\rm eV \AA}^6$,
$C_{71}^{\rm disp} = 4.93 \;{\rm eV \AA}^7$,
and $C_{73}^{\rm disp} = 1.19 \;{\rm eV \AA}^7$.

To find the effective potential of the impurity in the cluster, we will
once again integrate the above effective long range interaction
(Eq.~\ref{eq:V(r)}) over
the He atoms `missing' from the cluster.  
As before, we define the $z$ axis as along the
displacement of the molecule from the center of the cluster.  We
also define $\chi$ as the angle between symmetry axis of
the impurity and this $z$ axis.
In terms of these coordinates, we have the angle, $\phi$, between
this impurity symmetry axis and the vector from the impurity
to a point in space as:
\begin{equation}
\cos( \phi(\mbox{\boldmath $r$}') ) = \sin(\theta)\, \cos(\varphi)\, \sin(\chi) + 
\cos(\theta)\, \cos(\chi)
\end{equation}
where $\theta$ and $\varphi$ are the spherical coordinates of the point
{\boldmath $r$}$'$ defined relative to the position of the impurity.  
The results of the integration give the following results for the
effective potential for the impurity:
\begin{eqnarray}
\Delta E(a,R,\chi)  &=& V_2^0 F_2\,(a/R) +  V_4^0 F_4(a/R) P_2(\cos(\chi)\,) 
 + V_5^0 F_5(a/R) \cos(\chi) \nonumber \\
 &  & \,\,\, +  V_6^0 F_6(a/R) P_3(\cos(\chi)\,)  \label{eq:E_HCN} \\
V_4^0 & = & \frac{4 \pi \rho \, C_{62}}{3 R^3} \label{eq:V_4} \\
F_4(y) & = & \frac{1}{32} \left[ y\, (1 - y^2\,) \right]^{-3} 
\left[ -6y + 16 y^3 + 6 y^5 + 3(1-y^2\,)^3 \ln\left( \frac{1+y}{1-y} \right) 
\,\right]  \label{eq:F_4}\\
V_5^0 & = & \frac{4 \pi \rho \, C_{71}}{3 R^4} \label{eq:V_5} \\ 
F_5(y) & = & \frac{y}{ ( 1 - y^2 )^4 } \label{eq:F_5}\\
V_6^0 & = &  \frac{4 \pi \rho \, C_{73}}{3 R^4} \label{eq:V_6} \\
F_6(y) & = & \frac{15}{256} \left[ y ( 1 - y^2 ) \right]^{-4} \left[ 
  2 y^7 + \frac{146}{15} y^5 - \frac{22}{3} y^3 + 2 y \right. \nonumber \\
& & \;\;\; \left. 
- \ln \left( \frac{1+y}{1-y} \right) 
\left( y^8 - 4 y^6 + 6 y^4 - 4 y^2 + 1 \right) \right] \label{eq:F_6}
\end{eqnarray}
$F_2(y)$ and $V_2^0$ are unchanged from the previous definitions.
Despite the rather singular appearance of $F_4(y)$ and
$F_6(y)$ near $y \approx 0$, they are regular:
\begin{eqnarray}
F_4(y) & \approx & 
 \frac{3}{5}y^2 + \frac{12}{7}y^4 + \frac{10}{3} y^6 \ldots \\
F_5(y) & \approx & y + 4 y^3 + 10 y^5  \ldots \\
F_6(y) & \approx & \frac{15}{35} y^3  + \frac{15}{9} y^5 \dots 
\end{eqnarray}

\begin{figure}
\centerline{\epsfbox{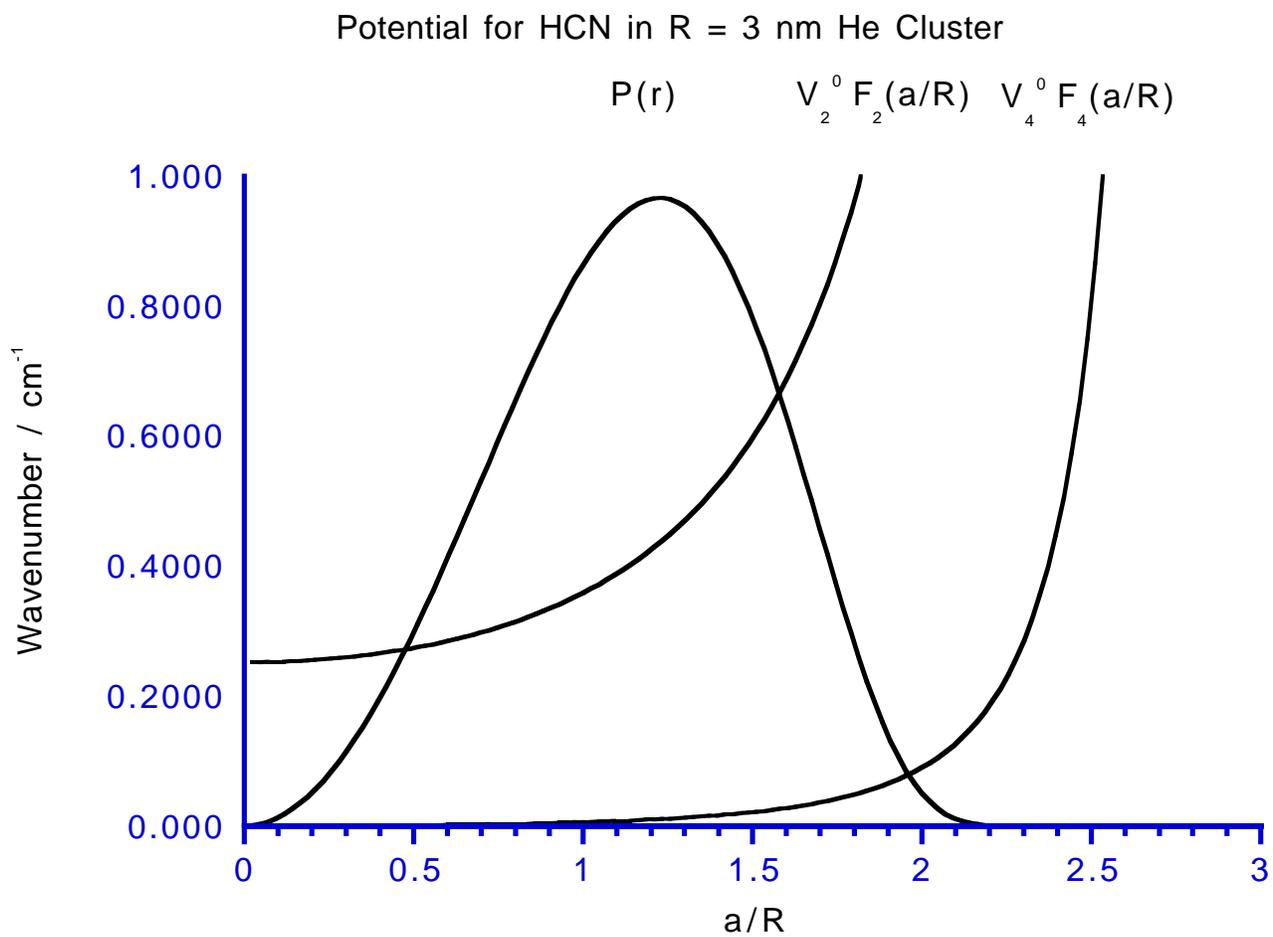}}
\vspace{1in}
\label{Potentials}
\caption{Comparison of isotropic ($P_0$) and anisotropic 
($P_2$) potentials for an HCN impurity in a Helium cluster of 3 nm radius. \hfill
}
\end{figure}

If we consider the case of HCN in a He cluster of $R = 3$\,nm, 
and use the parameters
given above, we get for the  effective potential:
\begin{eqnarray}
\Delta E(a,R,\chi) \frac{1}{hc} & = & (\, 0.2519 F_2(y)
 +0.0045 F_5(y) \cos(\chi)) \nonumber \\
& & \,\,\, +0.0619 F_4(y) P_2(\cos(\chi)) + 
0.0031 F_6(y) P_3(\cos(\chi)) \,)\, {\rm cm}^{-1}
\end{eqnarray}
We see that the two $C_7$ terms ($F_5(y)$ and $F_6(y)$)
can be neglected in first approximation 
compared to the isotropic and leading anisotropic term ($F_2(y)$ and $F_4(y)$) that
derive from the $C_6$ terms.
These coefficients are comparable to the line widths observed
in ro--vibrational spectra of molecules inside of He clusters.
The $V_2^0$ isotropic term will confine the impurity near the 
center of the cluster.  As discussed above, this
term determines the effective potential for the translational motion of the
the impurity, while the weaker anisotropic term will
couple the translational motion to the rotation of the
impurity molecule. Figure~\ref{Potentials} shows a plot of these two terms as a
function of $a$ for the case considered here, along with a plot
of the thermal probability density distribution as calculated from 
the isotropic potential, $V_2^0 F_2(y)$. It can be seen that the
anisotropic term, $V_4^0 F_4(y)$, is much smaller than the isotropic
term over the full range of displacements that are thermally well populated.

Since $C_{60} > 0$, it is lower in energy for the HCN axis to be perpendicular
to the displacement from the center of the cluster ($\phi = \pi/2$).  
This is easily rationalized if one considers the case of HCN below
a flat He surface.  When  $\phi = \pi/2$,
the nearest `missing Helium' is perpendicular to the HCN molecular axis, and
this cost less energy than when $\phi = 0$.  If the interaction energy is
much weaker than the seperation of rotational levels of the impurity,
which will be the case for HCN for all the thermally well populated
configurations, to first order we can consider the effective potential
for translational motion to be the average of the effective potential
over the rotational function of the impurity.  For a linear molecule,
the odd Legendre terms average to zero, while the $P_2$ term will split the
$m_j$ degeneracy of each $j$ level, such that the center of gravity is conserved.
If we consider the impurity center of mass displacement as static, then
the natural axis of quantization of the angular momentum is parallel to
this displacement, with an energy of a $j,m_j$ rotational 
state equal to:
\begin{equation}
\Delta E(a,J,M) = V_2^0 F_2\,(a/R) 
+  V_4^0 F_4(a/R) \left( \frac{j(j+1) - 3 m_j^2}{(2j-1)(2j+3)}\,\right)
\end{equation}
which implies that $m_j = 0$ states have the highest energy and $m_j = j$ the lowest
of each manifold of fixed $j$ states.

A proper quantum mechanical treatment of the coupling of the rotation
and translational motion of an anisotropic impurity will be given in the next
section.  This section will close by demonstrating a simple
`classical' application of the above results to estimate a line broadening
mechanism in the case of SF$_6$.  
The ground state of SF$_6$ has 
octahedral symmetry, and thus has only an isotropic 
long range interaction, as discussed above.
However, the IR allowed vibrational excitations are to triply degenerate
levels.  The $\nu_3$ mode observed in the IR spectrum of SF$_6$ in He
clusters has a particularly large vibrational transition dipole moment
of $\mu_3 =0.388$\,Debye for the $\nu_3 = 0 \rightarrow 1$ transition.
As the SF$_6$ is displaced in the $\nu_3$ mode, it will have a dipole
moment proportional to normal coordinate $Q_3$.  
This dipole result in a dipole--induced dipole interaction with the He, 
which modifies the force constant for
this normal mode~\cite{Eichenauer88}.  
The modification of the $\nu_3$ wavenumber
(relative once again to SF$_6$ in bulk liquid He), as a function of
the position of the SF$_6$ in the cluster and the orientation of the triply
degenerate $\nu_3$ mode are given by Equations~\ref{eq:Cind} and \ref{eq:E_HCN}, 
with the derivative of
the dipole with respect to the dimensionless normal coordinate
(which is just $\sqrt{2}$ larger than the fundamental transition dipole moment)
replacing the permanent dipole moment.  If we treat
these coordinates classically, we can calculate the spectrum of the 
$\nu_3$ mode of SF$_6$ in the cluster, relative to SF$_6$ in bulk liquid
He from the effective potential.   Thus, we can write $\nu_3(a,\chi)$ as
\begin{eqnarray}
\nu_3(a,R,\chi)  &=& \nu_3(\infty) \, + \, 
\nu^0 \left[  F_2\,(a/R) + F_4(a/R) P_2(\cos(\chi)\,) 
\,\right] \label{eq:nu_3} \\
\nu^0 & = & \frac{(\epsilon_r\, -1\,)\, 2 \mu_3^2}{12\, \pi\, \epsilon_r^2\, 
\epsilon_0\, R^3} 
\end{eqnarray}
$\nu_3(\infty)$ is the frequency of $\nu_3$ in the bulk He limit
and $\chi$ is the angle between the diplacement of the SF$_6$ and the
direction of polarization of the $\nu_3$ mode.
Note that this expression would be exact if $a$ did not change in time,
i.~e. we have a static inhomogeneous distribution of impurity positions,
as in a matrix without diffusion.  Dynamic motion of $a$ and $\chi$ will tend to
motionally average the instantaneous frequency $\nu_3(a(t),\chi(t))$ and
lead to a narrower spectrum than predicted by this approximation,
which is sometimes known as the 
`statistical method'~\cite{Townes55}
and was first introduced by Kuhn and London~\cite{Kuhn34} to
model pressure broadening.

\begin{figure}
\centerline{\epsfbox{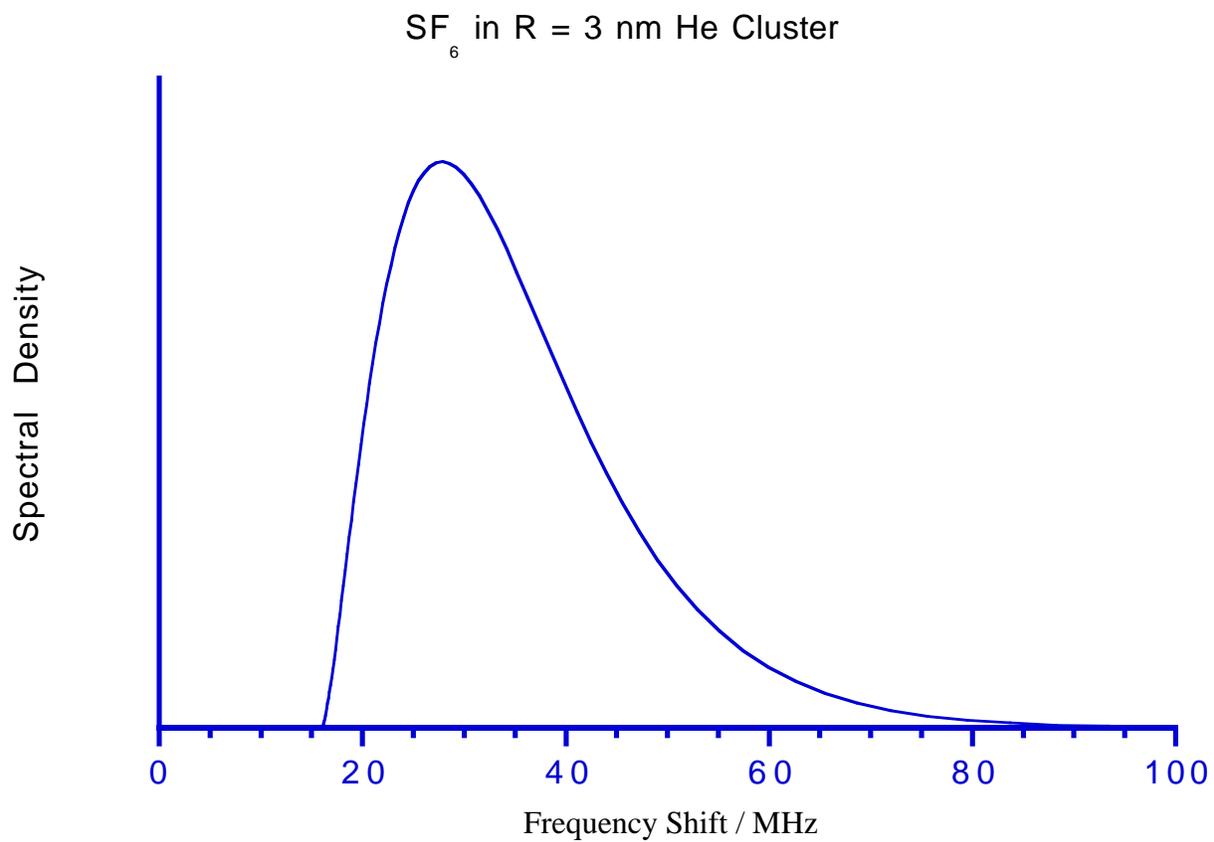}}
\vspace{1in}
\label{SF6_spec}
\caption{Calculated Spectral lineshape for the $\nu_3$ fundamental of SF$_6$
based upon the dipole-induced dipole spectral shift, which depends upon
the position inside the cluster
}
\end{figure}

For the case of the $\nu_3$ mode of SF$_6$ in a $R = 3$\,nm He
cluster, $\nu^0 = 16$\, MHz.
Since the isotropic term, $F_2(y)$ dominates over the
anisotropic term, $F_4(y)$, over the range of impurity displacements that
are thermally populated, we will keep only the isotropic term.
We can write the spectral density, $S(\nu)$ as
\begin{eqnarray}
S(\nu_3) &=& \left( \frac{d\nu_3}{da} \right)^{-1} \cdot
 Z_{\rm c}^{-1} \cdot 4 \pi a^2 \cdot  \exp (- V_2^0 F_2(y) /k_b T_{\rm c})  \label{eq:S} \\
   &=& \frac{24 \pi R}{Z_{\rm c}} \left[ 1 - \lambda^{-1} \right]^{3/2} \lambda^4 
\exp \left[ - \frac{V_2^0}{k_b T_c} \lambda^9 \right] \\
\lambda(\nu_3) & = & \left( \frac{\nu_3 - \nu(\infty)}{\nu^0} \right)^{1/3} 
\end{eqnarray}
where $ Z_{\rm c} = \int_0^R 4 \pi a^2 \exp (- V_2^0 F_2(a/R)/k_b T_{\rm c}) da$.
Figure~\ref{SF6_spec} shows the calculated spectral lineshape for $\nu_3$ of 
SF$_6$ in this model.  
It can be observed that the
residual shift of the mode (compared to
the limit of SF$_6$ in bulk He) and the
FWHM are both estimated to be about $30$\,MHz.
This can be compared to the linewidth of
$\approx 300$\,MHz FWHM observed by Hartmann {\it et al}~\cite{Hartmann96}. 
for the $\nu_3$ mode of SF$_6$ in $^4$He clusters of mean size 4000 atoms
($R = 3.5$\,nm).
It is clear that the calculated spectral 
broadening is more than an order of magnitude below that observed in the
laboratory, which argues that this effect contributes at most a
minor fraction of the total dephasing rate.

In their earlier paper that considered `particle in a box' states of
an impurity in He clusters, Toennies and Vilesov~\cite{Toennies95} proposed that
these levels could be directly observed as satellite lines on
a vibrational band of an impurity.  They suggested that
such particle in a box state changing transitions
could gain intensity due to differences in the effective potential in the
upper and low vibrational states, but reported no estimate of the expected
intensities of such transitions.  Using the SF$_6$ example, the author has
calculated Frank--Condon factors for transitions that change
the quantum numbers for the center of mass motion along with a vibrational
transition.  The largest such Frank--Condon factor was below
$\sim 10^{-4}$.  This is not unexpected since the excited state
has an effective potential that is stronger by only $\approx 0.1 \%$.
Based upon this analysis, it appears unlikely that such center of mass motion
changing transitions will be directly observed unless the change in
potential with vibrational coordinate is much larger.  It is worth pointing out
that the spectral structure that Toennies and Vilesov suggested may be due to such
transitions was latter assigned by them to the rotational structure of the 
vibrational transition inside the He cluster~\cite{Hartmann96}.

\section{Interaction of Impurity Translational and Rotational Motions}\label{sec-HCN}

Because the interaction between the impurity and the He cluster
depends upon both the position of the impurity in the cluster and
the orientational angles of the impurity, there is a coupling  that
can lead to relaxation of the orientational order, and thus line
broadening in the impurity spectra.  The purpose of this section is to 
elaborate how this can be calculated.  While the physical
origin of the effect is quite different, this problem has some
mathematical similarity to the coupling of
rotation and translation that has been used 
to model the IR spectra of light rotors
in solid noble--gas 
matrices~\cite{Redington62,Redington63,Friedmann65,Friedmann67}

We will assume our impurity is free to move in the cluster with an
effective mass $M_{\rm eff}$.  We will further assume it to be a linear molecule 
with an effective rotational constant $B_{\rm eff}$ in Helium.  We will model the 
system with the hamiltonian:
\begin{equation}
H = - \frac{\hbar^2}{2\,M_{\rm eff}} \nabla^2 + B_{\rm eff} \cdot 
\mbox{\boldmath $j$}^2 
+ V_2^0 F_2(y) + V_4^0 F_4(y) \cdot P_2(\cos(\theta_{12}))
\end{equation}
In this equation, $\nabla^2$ operates on the center of mass coordinates of the
impurity, {\boldmath $j$} is the angular momentum operator for
rotation of the impurity and $\theta_{12}$ is the angle between
the displacement of the impurity from the center of the cluster
and the direction of the linear axis of the impurity.
The results are easily extended to include other
Legendre terms, such as those that arise from the $r^{-7}$ terms
in the long range potential of  an isolated impurity and He atom.

If we keep all but the last term in our zero order hamiltonian,
then rotational and translational motion are separable and
we can write wavefunctions as:
\begin{equation}
|n,j,m_j,L,M_L> = \frac{\chi_{n,L}(a)}{a} \, Y_{j,m_j} (\Omega_1) \,
Y_{L,M_L} (\Omega_2)
\end{equation}
where $\chi_{n,L}(a)$ are the numerical solutions to the radial potential, $V_2^0 F_2(y)$,
with orbital angular momentum quantum number, $L$. 
$\Omega_1 = \theta_1,\phi_1$ are the
orientational angles of the rotor, and $\Omega_2 = \theta_2,\phi_2 $ the 
angles of the center of mass of the impurity,
but referenced to some arbitrary lab fixed axis system.
Introduction of the $V_4^0 F_4(y) P_2$ term will couple {\boldmath $j$ and $L$}, to
form a total angular momentum {\boldmath $J$}.  We
can write the basis function of the coupled representation as:
\begin{equation}
|n,j,L,J,M\,> = (-1)^{j - L} \sqrt{2 J + 1} 
\sum_{m_1,m_2} \left( \begin{array}{ccc}
j & L & J \\
m_1 & m_2 & -M
\end{array} \right) 
|n,j,m_1,L,m_2\,>
\end{equation}
The matrix elements of the zero order hamiltonian are diagonal in this 
representation as well, and give by:
\begin{equation}
E^0_{n,j,L} = E^0_{n,L} + B_{\rm eff} \cdot j(j+1)
\end{equation}
The $E^0_{n,L}$ and $\chi_{n,L}(a)$ can be determined by
numerical solution (using the Numerov--Cooley method) of the
radial Schr\"{o}dinger equation.

The coupling matrix elements can be calculated using standard
methods of angular momentum theory~\cite{Edmonds,Zare}, including the
Spherical Harmonic Addition Theorem to write
$P_2(\cos(\theta_{ij}))$ in terms of products of spherical harmonics
of coordinates $\Omega_1$ and $\Omega_2$:
\begin{eqnarray}
<n,j&&,L,J,M\,| H' | n',j',L',J,M> \, = V_4^0 \, <n,L\,| F_4(r) | n', L'\,>
 \cdot (-1)^{J+j+L'} \times \nonumber \\
&& \sqrt{(2j+1)(2j'+1)(2L+1)(2L'+1)}
  \cdot 
\left( \begin{array}{ccc}
j & j' & 2 \\
0 & 0 & 0
\end{array} \right) \cdot
\left( \begin{array}{ccc}
L & L' & 2 \\
0 & 0 & 0
\end{array} \right) \cdot
\left\{ \begin{array}{ccc}
J & j & L \\
2 & L' & j'
\end{array} \right\} 
\end{eqnarray}
The two $3J$ symbols in this expression lead to the 
selection rules for coupling that $j' = j, j \pm 2$
and $L' = L, L \pm 2$.
Explicit expressions for all the required $3J$ and $6J$
symbols can be found in the tables given by Edmonds~\cite{Edmonds}.
Since for the cases considered at present, the coupling matrix elements are comparable
to the separation of the center of mass motion states, but are
much less than the seperations of the rotational energy levels of the impurity,
the matrix elements that are off diagonal in $j$ are neglected in numerical calculations
presented below.  

The only change needed for other Legendre terms is to replace
the ``2'' in the above equations by the order of the Legendre
term, and the radial integral will be over the appropriate 
function, $F_n(y)$.  These terms will loosen the
selection rule on $\Delta j$ and $\Delta L$ for states that can be
coupled.  Parity, which is determined by $(-1)^{j+L}$, will remain
a good quantum number.

In order to calculate the dipole transition moments, we
will treat the electric field as along the laboratory Z axis.  This
gives:
\begin{equation}
\langle Y_{j,m_1} | \cos(\theta)\, | Y_{j',m_1} \rangle 
\, = \, \sqrt{(2j+1)(2j'+1)} \,
(-1)^{m_1} \, \left( 
\protect\ \begin{array}{ccc}
j & j' & 1 \\
0 & 0  & 0
\end{array} \protect\
\right) \cdot
\left( \begin{array}{ccc}
j    & j'  & 1 \\
-m_1 & m_1 & 0
\end{array}
\right)
\end{equation}
where $\theta$ is the angle between the Z axis and the symmetry axis of
the impurity molecule.
Going to the coupled representation we have:
\begin{eqnarray}
\langle n,j& &,L,J,M| \cos(\theta) | n',j',L',J',M' \rangle \, = \, 
(-1)^{j+j'+L+M+1}
\, \delta_{n,n'} \, \delta_{L,L'} \, \delta_{M,M'} \times \nonumber \\
& & \sqrt{(2j+1)(2j'+1)(2J+1)(2J'+1)} 
\left( \begin{array}{ccc}
j    & j'  & 1 \\
0 & 0 & 0
\end{array}
\right)
\left( \begin{array}{ccc}
J    & J'  & 1 \\
-M & M & 0
\end{array}
\right) \cdot
\left\{ \begin{array}{ccc}
L    & j'  & J' \\
1 & J & j
\end{array}
\right\}
\end{eqnarray}

We can expand the eigenfunctions in terms of the functions of
the coupled representation as:
\begin{equation}
| k, J, M > = \sum_{n,j,L} |n,j,L,J,M> <n,j,L,J | k, J>
\end{equation}
($k$ is a label to distinguish eigenstates with the same total angular 
momentum quantum numbers).
We do not put the $M$ label on the amplitudes since these are independent
of $M$ due to rotational invariance.
We can define a line strength factor for the intensity of a transition
between two levels $k,J$ and $k',J'$ as follows:
\begin{equation}
S(k,J; k',J') = 3 \sum_{M,M'} |<k, J, M | \cos(\theta) | k', J', M' > |^2 
\end{equation}
by exploiting the normalization of 3J symbols, we can reduce this to:
\begin{eqnarray}
\lefteqn{S(k,J; k',J') = (2J+1)(2J'+1) \times   } \nonumber \\
& & \left[ \sum_{n,j,j',L} <k,J|n,j,L,J><n,j',L,J'|k',J'> 
\sqrt{(2j+1)(2j'+1)} \right. \nonumber \\
& & \left. (-1)^{j+j'+ L} 
\left( \begin{array}{ccc}
j    & j'  & 1 \\
0 & 0 & 0
\end{array}
\right)
\left\{ \begin{array}{ccc}
L    & j'  & J' \\
1 & J & j
\end{array}
\right\} \right]^2 
\end{eqnarray}

\begin{figure}
\centerline{\epsfbox{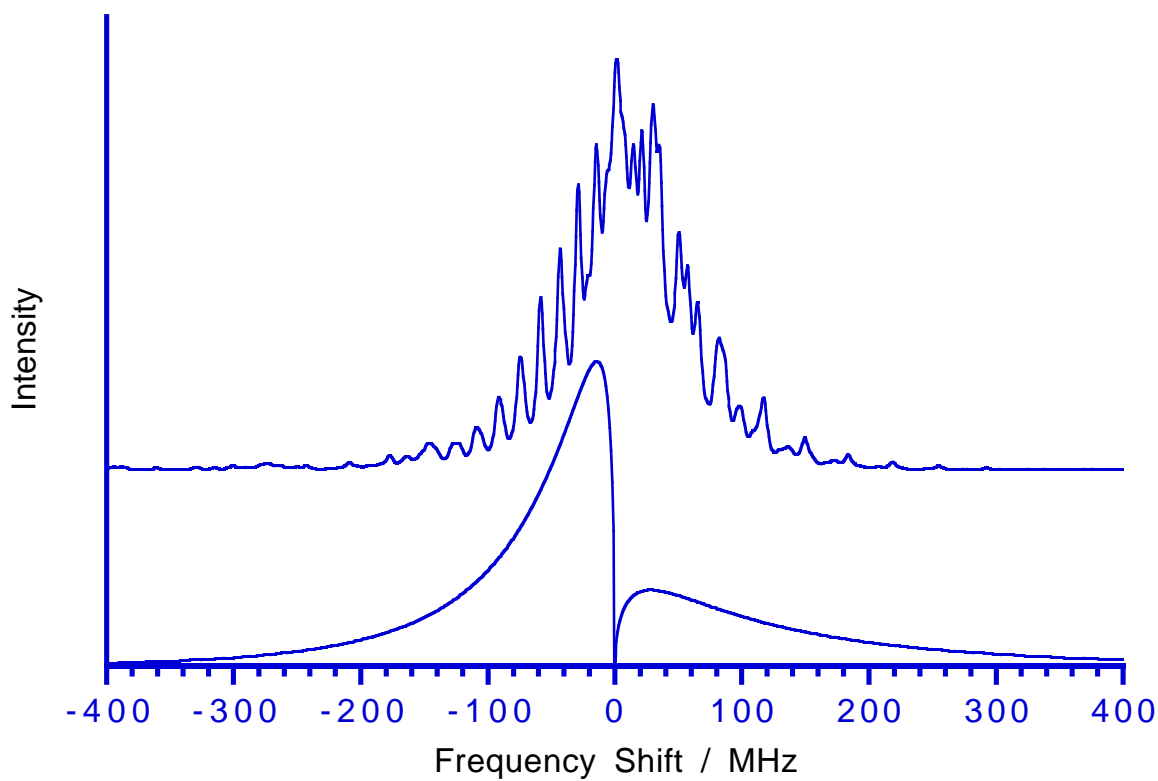}}
\vspace{1in}
\label{HCN_spec1}
\caption{Simulation of the HCN R(0) transition in a 3 nm radius Helium cluster, 
using the anisotropic potential discussed in the text, including
coupling between the HCN rotation and its center of mass motion relative to the
center of the He cluster.
}
\end{figure}

Figure~\ref{HCN_spec1} shows a simulation of 
the $j = 0 \rightarrow 1$ rotational
transition of HCN in a R = 3 nm He cluster.  The calculated `stick' spectrum
has been convoluted with a Lorentzian lineshape with FWHM
of $3$\,MHz.
A spectral broadening of the transition of $\approx 150$\,MHz 
can be seen in the figure.  The residual `fine' structure
would be washed out when one does an average over the experimental
cluster size distribution in a simulation of a real experiment, but
the overall FWHM of the transition is not a strong function of cluster size.
Also shown in the figure is the lineshape calculated from the semiclassical
`statistical' spectral model, where the spectrum is calculated as a Boltzman
average over different `static' center of mass positions for the HCN.  The 
two 'bumps' correspond to $J,M = 0,0 \rightarrow 1,\pm 1$ (more intense 
low frequency side) and $\rightarrow 1,0$ (broader and less intense
high frequency bump). These individual transitions are `washed out'
in the full quantum spectrum because the motion of the impurity causes
the natural quantization axis (the direction of the displacement from the
center of the cluster) to average over all directions.
The linewidth observed for the R(0) line in the first C--H overtone band
of HCN is $\approx 1.$\,GHz, an order of magnitude larger.  
This suggests that another mechanism dominates the dephasing rate in
the experiment.  

Recently, Nauta and Miller~\cite{Nauta99a}
have observed the R(0) transition in the C--H fundamental band
of HCN.  They find that the lineshape consists of two features,
with a width and seperation in qualitative agreement with the above
`statistical' model, i.e., as predicted without motional averaging.  
Furthermore, they have studied the spectrum as a function of
external electric field, and have found that the two features
correlate with the two high field 
$m = \pm 1$ and $m = 0$ peaks expected for the R(0) transition of
a polar linear molecule.

\section{Interaction of Impurity with He Cluster Excitations}\label{sec-Ripplons}

Up to now, we have considered the effect of the cluster potential on
the rotational and translational degrees of freedom of an impurity.
We will now consider the interaction of the impurity with the internal
vibrational excitations of the He cluster.  These can be separated into
two types.  The first are bulk excitations that consist of density
fluctuations or phonons.  The second are surface excitations that consist
of capillary waves and are known as ripplons.

As discussed by Brink and Stringari~\cite{Brink90}, 
for a cluster of $N$ atoms, the lowest
phonon excitation, the symmetric breathing mode, has a frequency of 
$540\, N^{-1/3}$ GHz.  Even for a cluster of $N = 10^4$ He atoms, this
corresponds to $25$\,GHz which is much larger than $k_b T_{\rm c} / h = 8.0$\,GHz.
Thus, only a few percent of such clusters have any quanta of excitation in
the phonon modes.   This implies that such modes cannot produce a
significant level of inhomogeneous broadening in impurity spectra.  
It is possible to observe excitations of the phonon modes in
an impurity absorption spectrum, but such excitations should
appear as a well separated `phonon side band' which peaks several 
$\rm cm^{-1}$ to the blue of the `zero phonon line' which does not involve
phonon excitations.  Such phonon wings are commonly seen in the 
electronic spectra of impurities in He~\cite{Stienkemeier95,Hartmann96a}, 
but have not yet been observed in
impurity vibration-rotation spectra.  This is presumably due to a 
small change in effective He solvation around impurities following
vibrational excitation.  As a result, almost all the transition
intensity is concentrated in the zero phonon line.

In contrast, the lowest frequency surface waves have much lower
wavenumbers, and excitations are thermally excited even at the low
temperature of He clusters.  These excitations involve no change in
He density and the only restoring force is due to the increase in
surface area caused by the change in the surface shape.  
The theory of such surface waves is presented in the
classic text on Hydrodynamics by H. Lamb~\cite{Lamb}, and discussed, in the
context of He clusters, by Brink and Stringari~\cite{Brink90}, 
who calculated the
spectra and thermodynamic functions of these excitations as a function
of cluster size.  

Due to excitation of ripplons, the surface of the He cluster develops oscillations.
If we consider a point of the surface of the cluster at polar angles
$\theta_s, \phi_s$ (again, relative to some arbitrary lab fixed
coordinate system), we can write the displacement of the surface as:
\begin{equation}
r(\theta_{\rm s},\varphi_{\rm s}) = R + \sum_{L_{\rm s},M_{\rm s}} 
S(L_{\rm s},M_{\rm s}) Y_{L_{\rm s},M_{\rm s}}
(\theta_{\rm s},\varphi_{\rm s})
\end{equation}
In the above equation, $S(L_{\rm s},M_{\rm s})$ is the normal coordinate of the 
surface wave with angular momentum quantum numbers
$L_{\rm s},M_{\rm s} $.  
The theory assumes that $S(L_{\rm s},M_{\rm s}) \ll R$.
The sum is for $L_{\rm s} \geq 2$.  The functions for
$L_{\rm s} = 0$ correspond to the breathing mode, and $L_{\rm s} = 1$ corresponds  to
a translation of the entire cluster, not distortions of the cluster shape. 
If we assume that the surface tension of He is 
$\sigma = 380\, {\rm \mu J m}^{-2}$~\cite{Brink90},
the frequency of the $L_{\rm s},M_{\rm s}$ surface mode is 
independent of $M_{\rm s}$ and is given by:
\begin{eqnarray}
\nu(L_{\rm s}) = \frac{\omega(L_{\rm s})}{2 \pi} 
&=& \frac{1}{2 \pi} \sqrt{L_{\rm s} (L_{\rm s} - 1) (L_{\rm s} + 2) 
\frac{\sigma}{M_{\rm He} \rho R^3}} \\
 &=&\, \sqrt{L_{\rm s} (L_{\rm s} - 1) (L_{\rm s} + 2) 
\frac{\sigma}{3 \pi M_{\rm He}}}\, N^{-1/2}
\nonumber \\
&=& 78\, {\rm GHz}\, \sqrt{L_{\rm s} (L_{\rm s} - 1) (L_{\rm s} + 2)} N^{-1/2} \nonumber
\end{eqnarray}
For a cluster of size $N = 10^4$, $\nu(L_s = 2) = 2.2$\, GHz.
The clusters will have a maximum 
value of $(L_{\rm s})_{\rm max} \approx (\pi/2) N^{1/3}$ which is estimated by assuming
that the smallest wavelength is $\approx$ twice the mean interparticle
distance.  For a cluster of $R = 3$\,nm, this corresponds to 
$(L_{\rm s})_{\rm max} = 26$.
Table 1 presents thermal average quantities for the ripplons
in clusters of sizes $R = 3 \; \& \; 5$\,nm, which can be compared with
the thermal averages of the impurity center of mass motion presented. 
Note that the average energy and entropy in the ripplons greatly exceeds
that of the impurity and forms the dominant `heat bath' of the
cluster.  However, the average thermal angular momentum in the impurity
center of mass motion approximately equals the total in the ripplons.
We can thus anticipate that in coming to equilibrium, the angular
momentum constraints may play an important role.  It is also worth
noting that for a cluster of size $R = 3$\,nm, the total canonical
internal energy at $T_{\rm c}$ is about equal to the energy required to
evaporate a single He atom, $\approx 18.8 k_b T_{\rm c}$~\cite{Brink90}.  
Thus, one can anticipate that smaller
clusters will be slightly warmer than larger clusters, due to their low heat capacity.
This is in contrast to the fact that we
expect a slight decrease in He binding energy for smaller clusters,
and thus higher vapor pressure, which would argue that smaller clusters
should be colder.  The decrease in binding energy, due to change in
surface free energy, can be estimated as 
$2 \sigma / (\rho R) = 1.73 \cdot {\rm cm}^{-1}\cdot {\rm nm}/R$, and is
considerably larger than the increase in binding energy caused by
attraction to a neutral impurity, which is 
$\approx 0.17\, {\rm cm}^{-1} \cdot$nm$^6 / R^6$ for the case of SF$_6$.

We can model the interaction between an impurity and a ripplon by
considering the change in the sum of impurity -- He interactions
caused by the modulation in the surface of the cluster.  Let 
$\delta r(\theta_{\rm s},\varphi_{\rm s})$ be the modulation in the cluster radius
in the direction $(\theta_{\rm s},\varphi_{\rm s})$ measured from the center of the
cluster.  (Note: this is a change from our previous treatments
where we used polar angles defined from the impurity).  
As above, let $y$ equal to the normalized distance of the impurity from the
center of the cluster.
To first order in $\delta r$, we can write the change in
He--impurity interactions as:
\begin{equation}
U(y) = \frac{C_{60} \rho}{R^4} \int \frac{\delta r(\Omega_{\rm s}) d\Omega_{\rm s}}
{\left[ 1 + y^2 -2 y \cos(\theta_{\rm is}) \right]^3 }+
\frac{C_{62} \rho}{R^4} \int \frac{\delta r(\Omega_{\rm s}) 
P_2(\cos(\theta_{\rm isr})d\Omega_s}
{\left[ 1 + y^2 -2 y \cos(\theta_{\rm is}) \right]^3 }
\end{equation}
where $C_{60}$ and $C_{62}$ are the isotropic and anisotropic contributions
of the $C_6$ coefficients, $\Omega_{\rm s}$ is the surface solid angle,
$\theta_{\rm is}$ is the angle between the vector from the center of the cluster
to the impurity
and the vector from the center of the cluster to the surface, and
$\theta_{\rm isr}$ is the angle between the vector from the impurity to the
point on the surface in direction $\theta_{\rm s}, \phi_{\rm s}$ from the
center of the cluster
and the vector along the symmetry axis of the impurity.
It is also useful to introduce the additional quantities as follows:
$r_{\rm is}$, the distance from
the impurity to the surface;
$\theta_{\rm ir}$, the angle between the vector from the center to the impurity
and the vector along the axis of the impurity;
and $\theta_{\rm sr}$, the angle between the vector from the center
to the surface and the vector along the axis of the impurity.
We can write these angles in terms of the three sets of polar angles
for the impurity center of mass ($\theta_i,\varphi_i$), 
the surface ($\theta_{\rm s},\varphi_{\rm s}$),
and the rotor ($\theta_{\rm r},\varphi_{\rm r}$) as follows:
\begin{eqnarray}
\cos(\theta_{\rm is}) &=& \cos(\theta_{\rm i}) \cdot \cos(\theta_{\rm s}) +
\sin(\theta_{\rm i}) \cdot \sin(\theta_{\rm s}) \cdot 
\cos(\varphi_{\rm i} - \varphi_{\rm s}) \\
\cos(\theta_{\rm ir}) &=& \cos(\theta_{\rm i}) \cdot \cos(\theta_{\rm r}) +
\sin(\theta_{\rm i}) \cdot \sin(\theta_{\rm r}) \cdot 
\cos(\varphi_{\rm i} - \varphi_{\rm r}) \\
r_{\rm is} &=& a \sqrt{ 1 + y^2 - 2 y \cos(\theta_{\rm is})} \\
\cos(\theta_{\rm isr}) &=& \frac{a}{r_{\rm is}} \left[\, \cos(\theta_{\rm sr}) - 
y \cos(\theta_{\rm ir}) \, \right] 
\end{eqnarray}

An exact treatment of this model would consider the simultaneous interaction
of the surface waves, the center of mass motion of the impurity, and
the rotation of the impurity, each of which has its own angular momenta
that must be coupled.  We will consider here instead a model for the 
interaction of the ripplons and impurity rotation alone.  As such, we
will treat the impurity as fixed at some normalized distance, $y$, from the center of
the cluster in a direction we will select for the $z$ axis.  In
the end, we will predict the spectrum by averaging over the possible 
positions $y$, using a Boltzmann weighting.  In this approximation we have
$\cos(\theta_{\rm is}\,) = \cos(\theta_{\rm s})$ and $\cos(\theta_{\rm ir}\,) = 
\cos(\theta_{\rm r})$.
In order to predict the $j = 0 \rightarrow 1$ rotational
transition of the impurity, we will need to average the
interaction over the lowest rotational functions of the rotor.  This
gives:
\begin{equation}
\langle Y_{00} (\Omega_r) \left| P_2 ( \cos(\theta_{\rm isr} ) 
\right| Y_{00} (\Omega_r) \rangle = 0 
\end{equation}
\begin{equation}
\langle Y_{10} (\Omega_r) \left| P_2 ( \cos(\theta_{\rm isr} ) 
\right| Y_{10} (\Omega_r) \rangle =
\frac{1}{5} \left[ \frac{  3 \cos^2 (\theta_{\rm s}) - 4 y \cos(\theta_{\rm s}) 
+ 2 y^2 - 1}{ 1 + y^2 - 2 y \cos(\theta_{\rm s}) } \right] 
\end{equation}
\begin{equation}
\langle Y_{1 \pm1} (\Omega_{\rm r}) \left| P_2 ( \cos(\theta_{\rm isr} ) 
\right| Y_{1 \pm1} (\Omega_{\rm r}) \rangle =
-\frac{1}{2} <Y_{10} (\Omega_{\rm r}) | P_2 ( \cos(\theta_{\rm isr} ) |
 Y_{10} (\Omega_{\rm r}) >
\end{equation}
We thus find that the $j = 0$ rotor state does not couple to the ripplons,
while the $j = 1 $ state does.

In order to calculate matrix elements of the interaction, we need to
determine the dimensionless amplitude for the ripplon vibrations.
Using the methods presented in Lamb's text~\cite{Lamb}, we can determine the 
effective mass, $\mu(L_{\rm s})$, for each ripplon:
\begin{equation}
\mu(L_{\rm s}) = \frac{ M_{\rm He} \, \rho \, R^3}{L_{\rm s}}
\end{equation}
Note that the effective mass is proportional to the total mass of the cluster,
not the mass of atoms near the surface. This is due to the fact that the 
surface waves have a velocity field that goes into the bulk of the
cluster, with the velocity proportional to $y^{L_{\rm s}-1}$.
Surface waves of higher $L_{\rm s}$ values are increasingly confined to 
motion of atoms
closer to the surface, and thus have a reduced effective mass.
Using this relationship, we can write the surface displacement in
terms of raising and lowering operators for the ripplons 
($a^{\dagger}(L_{\rm s},M_{\rm s})$ and $a(L_{\rm s},M_{\rm s})$) as follows:
\begin{eqnarray}
\delta r(\theta_{\rm s},\varphi_{\rm s}) &=& \sum_{L_{\rm s},M_{\rm s}} 
S(L_{\rm s},M_{\rm s}) Y_{L_{\rm s},M_{\rm s}}
(\theta_{\rm s},\varphi_{\rm s}) \\
S(L_{\rm s},M_{\rm s}) &=& 
d_{\rm s}( L_{\rm s} )  
\left(\, a^{\dagger}(L_{\rm s},M_{\rm s}) + a(L_{\rm s},M_{\rm s}) \, \right) \\
d_{\rm s}( L_{\rm s} ) &=& \sqrt{\frac{\hbar}{2 \, \mu(L_{\rm s}) \, \omega(L_{\rm s}) }} =
\left[ \frac{L_{\rm s} \hbar^2}{2 (L_{\rm s}-1) 
(L_{\rm s} + 2) \, \sigma \, M_{\rm He} \, \rho R^3 \,}
\right]^{1/4}
\end{eqnarray}
For $R = 3$\, nm, $d_{\rm s}( 2) = 0.37 {\rm \AA}$.  
Summing the thermally averaged displacement of all ripplons up to $(L_{\rm s})_{\rm max}$
gives an RMS thermal motion of the surface of $1.34 (1.50)$\,\AA\ for
clusters of $R = 3(5)$\,nm radii.  

We can thus write the interaction of ripplons with the rotation of the
impurity molecule as follows:
\begin{eqnarray}
{\cal H}_{\rm ripplon,rotation} &=& U_0 
 \sum_{L_{\rm s},M_{\rm s}} 
\left( a^{\dagger}(L_{\rm s},M_{\rm s}) + a(L_{\rm s},M_{\rm s}) \right)
\times \nonumber \\
&  & \left( \frac{L_{\rm s}}{(L_{\rm s} - 1)(L_{\rm s} + 2)} \right)^{1/4}
\int \frac{Y_{L_{\rm s},M_{\rm s}} (\Omega_{\rm s}) 
P_2 (\cos( \theta_{\rm isr}\,)\,) d\Omega_{\rm s}}
{\left[ 1 + y^2 - 2 y \cos(\theta_{\rm is}) \right]^3 } \label{eq:U}
\end{eqnarray}
\begin{equation}
U_0 = \left( \frac{C_{60}\,\rho}{R^4} \right) 
\left[ \frac{ \hbar^2}
{2 \, \sigma \, M_{\rm He} \, \rho R^3 \,}
\right]^{1/4}
\end{equation} 
For an HCN impurity in a 5 nm He cluster, $U_0 = 1.3\,{\rm MHz} \cdot h$
 
Despite the fact that the integral diverges as $y \rightarrow 1$, the
coupling matrix elements diagonal in rotational quantum numbers are
much less than the spacing between ripplon states for the range of
$y$ that includes most of the probability density of the impurity.
Since the coupling is linear in the ripplon coordinates (which are
harmonic oscillators), ${\cal H}_{\rm ripplon,rotation}$ has 
no first order contribution to the energy. 
This suggests that we can treat this coupling by second order perturbation
theory to account for the virtual creation and annihilation of ripplons as
an impurity rotates.  A consequence of the linear
ripplon coupling is that this second order correction is
independent of the occupation of the ripplon modes!
Such coupling shifts the equilibrium position of the ripplons from
zero displacement but does not change their force constant.  
We find a shift in energy equal to:
\begin{eqnarray}
\Delta E (y, j, m_j) &=& 
\sum_{L_{\rm s},M_{\rm s}} - \frac{U_0^2}{h \nu (L_{\rm s})}
\left( \frac{L_{\rm s}}{(L_{\rm s} - 1)(L_{\rm s} + 2)} \right)^{1/2} 
\times \nonumber \\
&  &  \left[ 
\int \frac{Y_{L_{\rm s},M_{\rm s}} (\Omega_{\rm s}) 
\langle Y_{j, m_j} (\Omega_{\rm r})\, 
|\, P_2 (\cos( \theta_{\rm isr}\,)\,) | \,
 Y_{j,m_j} (\Omega_{\rm r})  \rangle d\Omega_{\rm s}}
{\left[ 1 + y^2 - 2 y \cos(\theta_{\rm is}) \right]^3 }  \right]^2
\end{eqnarray}
If we select the z axis as along the vector from the center of the
cluster to the impurity center of mass, then only the $M_{\rm s} = 0$
terms contribute to the sum.

\begin{figure}
\centerline{\epsfbox{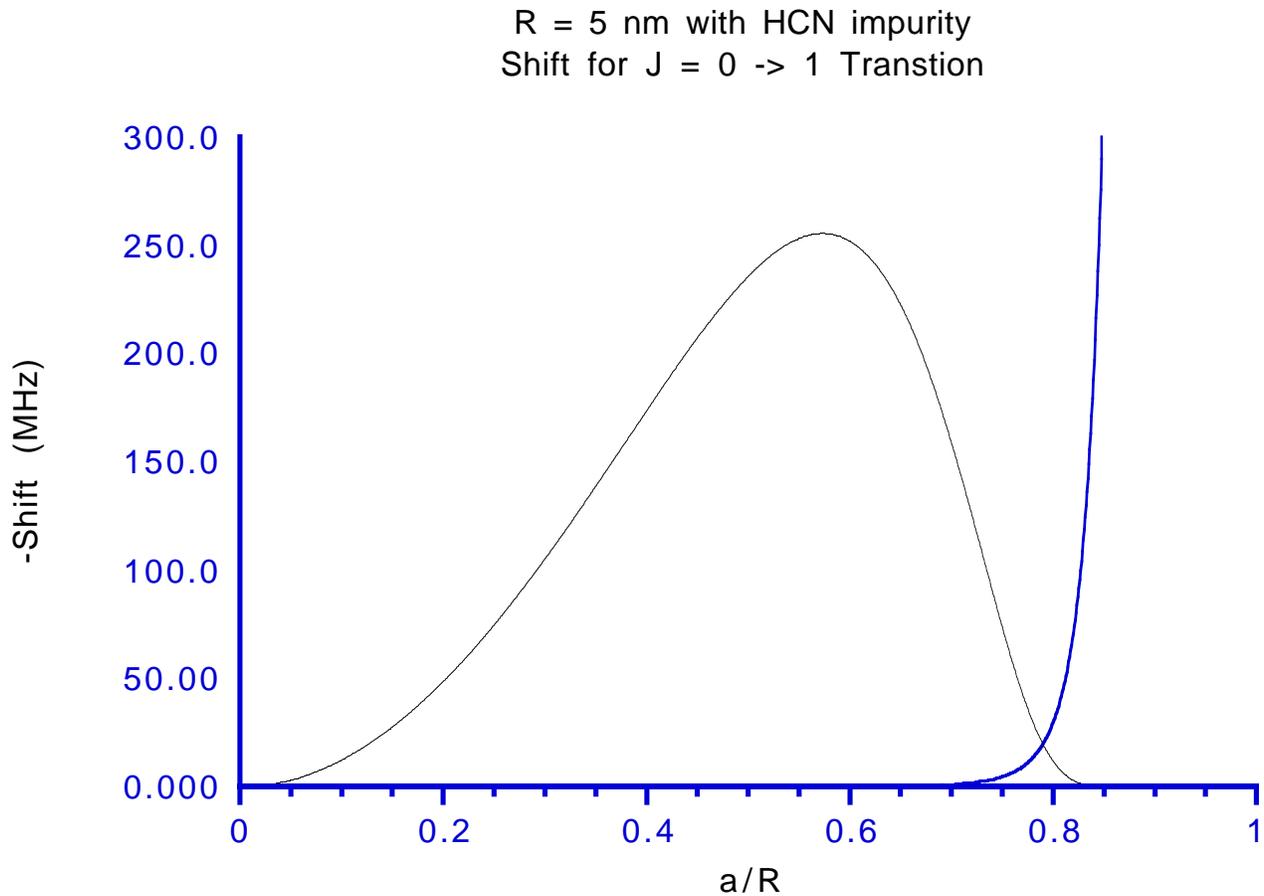}}
\vspace{1in}
\label{Ripplon_shifts}
\caption{Calculated shift in HCN $j = 1, m_j = 0$ level as a result of the
interaction with the surface ripplons,
as a function of impurity displacement from the center of the cluster.
Also plotted is the radial distribution function in arbitrary units to
show region of thermal population.
}
\end{figure}

The above expression was used to calculate the shift in the energy of
the $j = 1, m_j = 0$ level as a function of $y$.  
The integrals over $\theta_{\rm s}$ were performed numerically
using Gauss--Legendre integration.  The shift for the
two $j = 1, m_j = \pm 1$ levels will be just one quarter as large.  
Figure~\ref{Ripplon_shifts} shows the 
calculated shift divided by Planck's
constant as a function of $a$, along with a plot of the radial distribution function
for the impurity.  This was calculated from the isotropic, $C_{60}$, interaction
energy, as described above.  It can be seen that the ripplon induced
shift is completely negligible except for the extreme tail of the 
Boltzmann distribution.  We can conclude that the potential interaction of
the impurity rotation with the ripplons can be neglected compared to
the position dependent anisotropic dispersion interaction.

\section{Hydrodynamic Coupling of Impurity motions}\label{sec-Hydro}

While the motions of the impurity in the superfluid He are expected
to be without friction, the medium does exert an influence by
hydrodynamic effects.  The reason is that as the impurity
rotates or translates, He atoms must move out of the way to avoid the
short range repulsion.
For a spherical impurity, the effect of the He kinetic energy
produced by the impurity translational motion can be modeled by
an effective mass for the impurity that is increased by
0.5 times the mass of He displaced by the impurity.
For an anisotropic impurity, there is in addition both
a hydrodynamic contribution to the effective moment of inertia
for rotation, and a coupling between rotation and center of mass
motion.  In this section a detailed calculation of these
effects will be presented.  
This type of coupling was previously discussed (though developed
in much less detail) in a paper by Elser and Platzman, who proposed
an anisotropic solvation structure around positive ions in
bulk liquid He to explain an observed temperature dependence to the effective
mass of ions~\cite{Elser88}.  

The hydrodynamic coupling of translation and rotation does not
occur for an impurity with tetrahedral or octahedral symmetry~\cite{Lamb}.
Let us model the impurity molecule as a rigid, symmetric ellipsoid of principal 
axes $a = b = 3.2 {\rm \AA}$
and $c = \xi a = 3.7 {\rm \AA}$.  These parameters are based upon the position
of the inner wall in the He--HCN potential of Atkins and Hutson~\cite{Atkins96}.
Since the dominant hydrodynamic effect
is for parts of the fluid close to the moving body, the use of a
continuum, incompressible fluid model is a questionable approximation.
However, because of its simplicity, it is worth working out in detail
the predictions of this model.  Further, it may be the case that the 
form of the resulting hamiltonian may be more generally applicable than
the simple hydrodynamic model used to derive it.

\begin{figure}
\centerline{\epsfbox{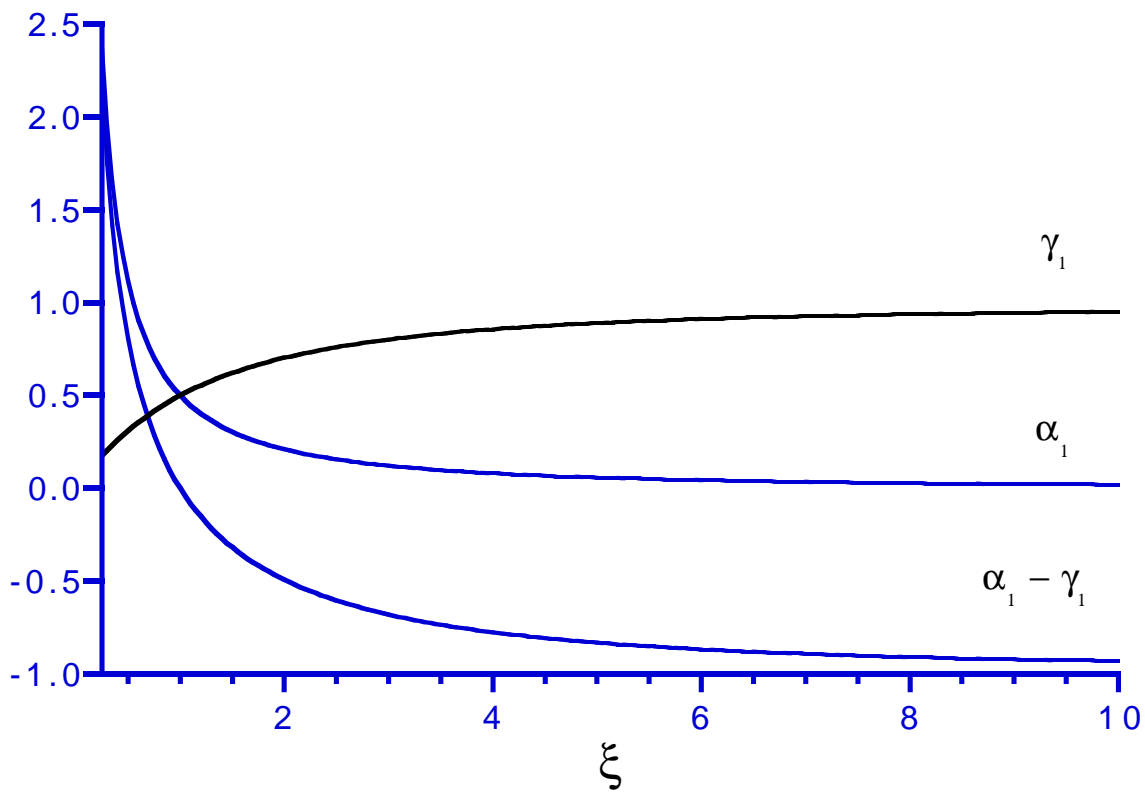}}
\vspace{1in}
\label{Hydro}
\caption{The hydrodynamic constants $\alpha_1$ and $\gamma_1$ and their
difference as a function of $\xi$, which is the ratio of the size
of the ellipsoid along the symmetry axis divided by the size perpendicular
to this. 
}
\end{figure}

The motion of an ellipsoid in an incompressible fluid without 
viscosity is worked out in the text by Milne-Thomson~\cite{Milne}.
The hydrodynamic kinetic energy is given for an ellipsoid
moving in any direction relative to its principal axes.  
$\vec{v}$ denotes the velocity of the ellipsoid relative to
the fluid at large distance from the ellipsoid;  
$\vec{\lambda}$ denotes a unit vector in the direction of the
unique principal axis; $M' = M_{He}\, \rho\, \frac{4}{3}\, \pi\, a^3\, \xi$
is the mass of He displaced by the ellipsoid ($34.6$\, u for
the HCN--He parameters given above), and M the mass of
the ellipsoid.  The classical kinetic energy of translational motion
of the ellipsoid is:
\begin{equation}
T = \frac{1}{2} \left( \,M + \frac{1}{3} M' (2\gamma_1 + \alpha_1) \right)
 \vec{v} \cdot \vec{v} + \frac{1}{2} M' (\alpha_1 - \gamma_1 )
\left[ \left( \vec{\lambda} \cdot \vec{v} \right)^2 - \frac{1}{3} 
\vec{v} \cdot \vec{v} \right]
\end{equation}
where
\begin{eqnarray}
\alpha_1 = \frac{\alpha_0}{2-\alpha_0} \;\;\; &   &  \;\;\;
\gamma_1 = \frac{\gamma_0}{2 - \gamma_0} \\
\alpha_0 = \int_0^{\infty} \frac{\xi dx}{(1+x) (\xi^2 + x)^{3/2}} \;\;\; &   &
\;\;\; \gamma_0 = \int_0^{\infty} \frac{\xi dx}{(1+x)^2 \sqrt{\xi^2 + x}}
\label{eq:alpha_0}
\end{eqnarray}
Figure~\ref{Hydro} shows $\alpha_1$, $\gamma_1$, and their difference
as a function of $\xi$.  We can see that the coupling between
translational motion and rotation is zero in the case of a sphere ($\xi = 1$),
saturates for a very prolate ellipsoid ($\xi \gg 1$), and diverges for
an extremely oblate ellipsoid ($\xi \rightarrow 0$), i.e. for a flat
disk.  
For our model of HCN in He, $\xi = 1.16$, and we have
$\alpha_1 = 0.419   $ and $\gamma_1 = 0.544  $.
The coupling of $\vec{\lambda}$ and $\vec{v}$ reflects the fact that
it takes less energy (at fixed velocity) to move the ellipsoid though the fluid with
its long principal axis aligned with the flow, then with this
axis perpendicular to the flow.  This leads to coupling of the
rotational motion to the translational motion.  In the bulk,
the rotational axis will precess about the velocity vector, and this
would be expected to produce re--orientational dephasing of a molecular
ro--vibrational absorption line, since there is a distribution of 
molecular translational velocity.

In order to derive the quantum hamiltonian for the motion of such
an ellipsoid, we need to re--express the kinetic energy in terms of 
the momentum of the ellipsoid, defined by:
\begin{equation}
\vec{p} = \nabla_{v} T = \left[ M + M' \gamma_1 \right] \vec{v}
+ M' \left( \alpha_1 - \gamma_1 \right) 
\left( \vec{\lambda} \cdot \vec{v} \right) \vec{\lambda}
\end{equation}
which can be inverted to write
\begin{equation}
\vec{v} = \frac{1}{M + M' \gamma_1} \left[ \vec{p}
- \frac{M' (\alpha_1 - \gamma_1) (\vec{\lambda} \cdot \vec{p} )}
{M + M' \alpha_1} \vec{\lambda} \right]
\end{equation}
Substituting into the above expression for the kinetic energy gives
\begin{equation}
T = \frac{1}{2 \left( M + M' \gamma_1 \right)} \vec{p} \cdot \vec{p}
- \frac{M' \left( \alpha_1 - \gamma_1 \right)}
{2 \left(M + M' \alpha_1 \right) \left( M + M' \gamma_1 \right) }
\left( \vec{\lambda} \cdot \vec{p} \right)^2
\end{equation}
Making the substitution $ \vec{p} \rightarrow i \hbar \nabla$ we
get the quantum hamiltonian:
\begin{equation}
{\cal H} = -\frac{\hbar^2}{2 M_{\rm eff}} \nabla^2 + 
C_{\rm H} \left[ \left( \vec{\lambda} \cdot \vec{\nabla} \right)^2
-\frac{1}{3} \nabla^2 \right]
+ B_{\rm eff} \mbox{\boldmath $J$}^2 + V(\vec{r},\vec{\lambda})
\end{equation}
where
\begin{equation}
M_{\rm eff} = \frac{3(M+M'\alpha_1)(M + M' \gamma_1)}
{3 M + 2 M' \alpha_1 + M' \gamma_1}
\end{equation}
\begin{equation}
C_{\rm H} = \frac{\hbar^2 \, M' \left( \alpha_1 - \gamma_1 \right)}
{2 \left(M + M' \alpha_1 \right) \left( M + M' \gamma_1 \right) }
\end{equation}
For our model of HCN in He, $C_{\rm H}/h = -0.772\, {\rm GHz \, \AA}^2$, and
$M_{\rm eff} = 35.2$\,u. 

\begin{figure}
\centerline{\epsfbox{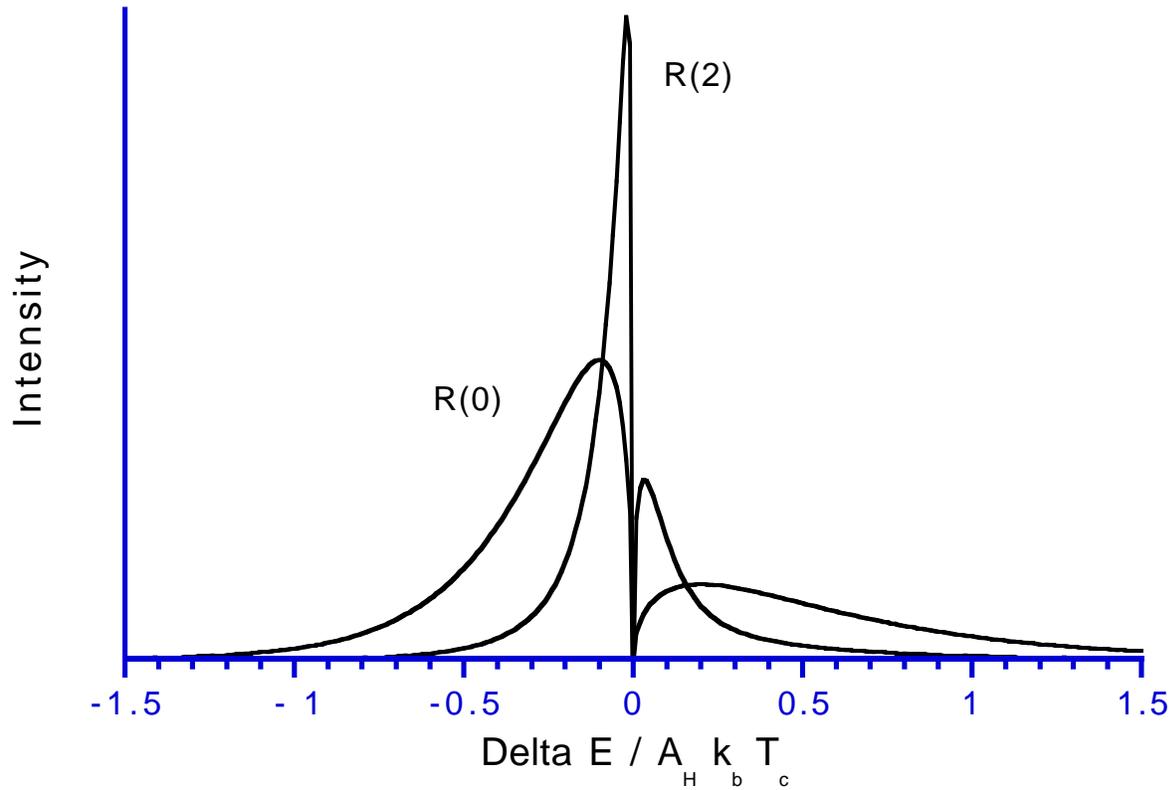}}
\vspace{1in}
\label{Bulk_He_Spec}
\caption{Calculated Spectrum for the R(0) and R(2) lines of a
linear molecule in bulk liquid He.  The ordinate is plotted in units of
$A_{\rm H} k_b T_{\rm c} h^{-1}$, where $A_{\rm H}$ is defined in the text,
$T_{\rm c}$ the temperature of the liquid He, and the rest are standard symbols
for physical constants. }
\end{figure}

Before considering the case of an impurity in a cluster, it is
instructive to examine the case of a linear molecule impurity in bulk
He, where the momentum is a good quantum number.  If we assume that
we can treat the hydrodynamic coupling by first order perturbation theory,
the shift in energy for a level $j,m_j$ (with the quantization axis along
the velocity), and translational kinetic energy $E_{\rm trans}$, is given by:
\begin{eqnarray}
E_{\rm hydro}(j,m_j) & = & A_{\rm H} E_{\rm trans} \frac{j(j+1) - 3 m_j^2}{(2j-1)(2j+3)} \\
A_{\rm H}             & = & (-1) \left( \frac{2 M' ( \alpha_1 - \gamma_1 )}
{3 M + 2 M' \alpha_1 + M' \gamma_1} \right)  \label{eq:A_H}
\end{eqnarray}
For the case of HCN, $A_{\rm H} = 0.036$.  The $j = 0 \rightarrow 1$
transition will be shifted by $+0.4 A_{\rm H} E_{\rm trans}$
and $-0.2 A_{\rm H} E_{\rm trans}$
for polarization parallel and perpendicular to the velocity vector
of the impurity.
The width of a $j \rightarrow j+1$ transition will scale as $1/j$.
Figure~\ref{Bulk_He_Spec} shows the calculated spectral lineshapes for
the R(0) and R(2) transitions after averaging over the Maxwell--Boltzmann
distribution for $E_{\rm trans}$.
The horizontal axis is in units of
$A_{\rm H} k_b T_{\rm c} h^{-1}$ which equals 283 MHz for the case of HCN.
The spectra have a cusp in the center due to the $\sqrt{E_{\rm trans}}$ factor 
in the density of states. The higher frequency lobe arises from parallel
transitions and the lower lobe, the perpendicular transitions
(relative to the impurity momentum vector).
Note that molecules with the projection of the rotational angular momentum 
perpendicular to the center of mass momentum are higher in energy than
those with rotational angular momentum either parallel or antiparallel to
the linear momentum.  This is because in the constant momentum case, orientations
with the linear molecule axis parallel to the linear momentum are higher in
energy, due to the lower effective mass for that orientation.   

Explicit expressions for the matrix elements
of the hydrodynamic term can be derived from the relationships from
the matrix elements of $\vec{\lambda}\cdot\vec{\nabla}$
operating on the spherical harmonics, as given by Rose~\cite{Rose54}. 
After {\it considerable}
algebra, the matrix elements diagonal in $j$ in the coupled representation
(where $L$ is the center of mass motion orbital angular momentum quantum number, 
$j$ the quantum number of rotation of the linear impurity, and 
$J$ the quantum number for the total angular momentum of the impurity),  
\begin{equation}
| n,L,j, J, M > \, =  \sum_{m_1,m_2} (-1)^{L - j + M} \,
\sqrt{2 J + 1} \,
\left( \begin{array}{ccc}
L    & j  & J \\
m_1 & m_2 & -M
\end{array} \right) 
\frac{\chi_{n,L}}{r} \, Y_{L,m_1}(\Omega)\,  Y_{j,m_2}(\Omega_r) 
\end{equation}
can be found.  For $L' = L + 2$ they are:
\begin{eqnarray}
\langle n',&  & L+2,j,J, M | 
\left( \vec{\lambda} \cdot \vec{\nabla} \right)^2 -\frac{1}{3}\nabla^2
| n, L, j, J, M \rangle \, = \, (-1) \sqrt{(L+1)(L+2)} \times \nonumber \\
&  &   \left[ (j+1) 
{ \left\{ \begin{array}{ccc}
J & L     & j \\
1 & j+1 & L+1
\end{array} \right\}
\left\{ \begin{array}{ccc}
J & L+1   & j+1 \\
1 & j & L+2
\end{array} \right\} } \right. \nonumber \\
& & \;\;\;\; \left.
+ j
{ \left\{ \begin{array}{ccc}
J & L     & j \\
1 & j-1 & L+1
\end{array} \right\}
\left\{ \begin{array}{ccc}
J & L+1  & j-1 \\
1 & j  & L+2
\end{array} \right\} }
\right] \times \nonumber \\
&  & \int_0^R \chi_{n',L+2}(r) \left[
\left( \frac{2 M_{\rm eff}}{\hbar^2} \right) 
\left( V(r) - E_{n,L}^0 \right)
+ \frac{(L+1)(2L+3)}{r^2} - \frac{2L+3}{r} \frac{d}{dr} \right] \chi_{n,L}(r) dr 
\end{eqnarray}
For $L' = L-2$ they are:
\begin{eqnarray}
\langle n',&  & L-2,j,J, M | 
\left( \vec{\lambda} \cdot \vec{\nabla} \right)^2 -\frac{1}{3}\nabla^2
| n, L, j, J, M \rangle \, = \, (-1) \sqrt{(L(L-1)} \times \nonumber \\
&  &   \left[ (j+1) 
{ \left\{ \begin{array}{ccc}
J & L     & j \\
1 & j+1 & L-1
\end{array} \right\}
\left\{ \begin{array}{ccc}
J & L-1   & j+1 \\
1 & j & L-2
\end{array} \right\} } \right. \nonumber \\
& & \;\;\;\; \left.
+ j
{ \left\{ \begin{array}{ccc}
J & L     & j \\
1 & j-1 & L-1
\end{array} \right\}
\left\{ \begin{array}{ccc}
J & L-1  & j-1 \\
1 & j  & L-2
\end{array} \right\} }
\right] \times \nonumber \\
&  & \int_0^R \chi_{n',L-2}(r) \left[
\left( \frac{2 M_{\rm eff}}{\hbar^2} \right) 
\left( V(r) - E_{n,L}^0 \right)
+ \frac{L(2L-1)}{r^2} + \frac{2L-1}{r} \frac{d}{dr} \right] \chi_{n,L}(r) dr 
\end{eqnarray}
And for the case $L' = L$ they are:
\begin{eqnarray}
\langle n',&  & L,j,J, M | 
\left( \vec{\lambda} \cdot \vec{\nabla} \right)^2 -\frac{1}{3}\nabla^2
| n, L, j, J, M \rangle \, = \,  \nonumber \\
&  &   \left[ (L+1)(j+1) 
\left| \left\{ \begin{array}{ccc}
J & L     & j \\
1 & j+1 & L+1
\end{array} \right\} \right|^2
+(L+1)j 
\left| \left\{ \begin{array}{ccc}
J & L   & j \\
1 & j-1 & L+1
\end{array} \right\} \right|^2 \right. \nonumber \\
& & \;\;\;\; \left.
+ L (j + 1)
\left| \left\{ \begin{array}{ccc}
J & L   & j \\
1 & j+1 & L-1
\end{array} \right\} \right|^2
+L j
\left| \left\{ \begin{array}{ccc}
J & L   & j \\
1 & j-1 & L-1
\end{array} \right\} \right|^2 -\frac{1}{3}
\right] \times \nonumber \\
&  & \int_0^R \chi_{n',L}(r) \left[
\left( \frac{2 M_{\rm eff}}{\hbar^2} \right) 
\left( V(r) - E_{n,L}^0 \right)
 \right] \chi_{n,L}(r) dr 
\end{eqnarray}
For the case $j = 0$, $J = L = L'$, and the matrix element 
reduces to:
\begin{equation}
\langle n', L,0,L, M | 
\left( \vec{\lambda} \cdot \vec{\nabla} \right)^2 -\frac{1}{3}\nabla^2
| n, L, 0, L, M \rangle \, = 0.
\end{equation}
which is the reason that the $-\frac{1}{3}\nabla^2$ term was
included as part of the hydrodynamic term.
$V(r)$ is the isotropic part of $V\left(\vec{r},\vec{\lambda}\right)$
and the basis states are assumed to be eigenstates of 
 ${\cal H}_0$ defined by:
\begin{equation}
{\cal H}_0 = - \frac{\hbar^2}{2 M_{\rm eff}} \nabla^2 + V(r)
\end{equation}
with eigenvalues $E_{n,L}^0$.
While not obvious by inspection, the above matrix elements can be
shown to describe a Hermitian operator, as they should.

\begin{figure}
\centerline{\epsfbox{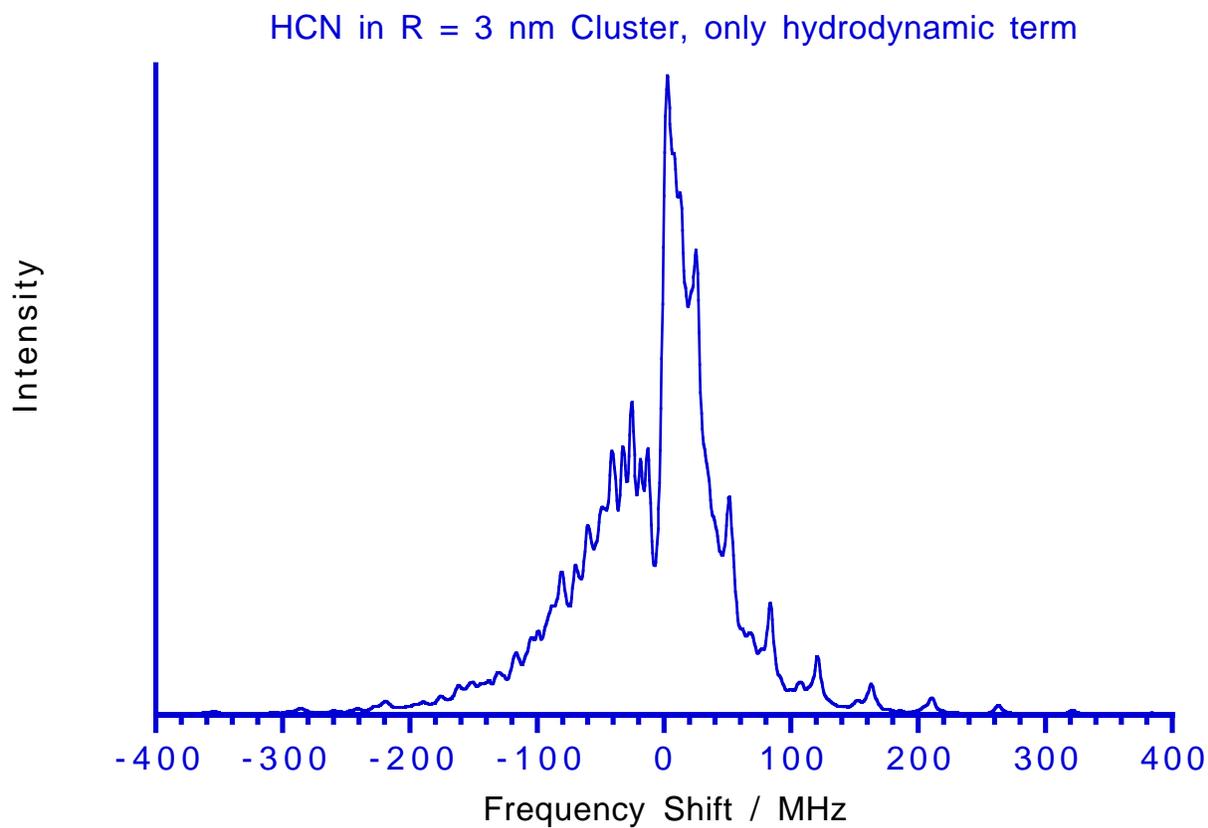}}
\vspace{1in}
\label{HCN_spec2}
\caption{Calculated HCN R(0) line shape, including the isotropic
term in the potential and the hydrodynamic coupling, for a He cluster
of $R = 3$\,nm.}
\end{figure}

Using this hamiltonian, the $j = 0 \rightarrow 1$ spectrum
of HCN in a He cluster can be recalculated. 
Figure~\ref{HCN_spec2} shows the spectrum calculated including the
isotropic ($C_{60}$) term in the potential and the hydrodynamic coupling,
but without the anisotropic potential term ($C_{62}$) for a cluster of
$R = 3$\,nm.  The calculated spectrum of the impurity in the cluster
looks remarkably like that (see Figure~\ref{Bulk_He_Spec}) for a 
linear molecule impurity in bulk liquid He, except the spectrum is `inverted'. 
This can be understood as a consequence of averaging over the 
classical trajectory of the motion, which is in a plane perpendicular
to $L$.  For circular orbits, this leads to the prediction that the 
$j = 1$ state with $m_j = 0$ is
higher in energy than the states with $m_j = \pm 1$ by an amount
equal to $2 (-C_{\rm H}) M_{\rm eff} E_{\rm k} / 5 \hbar^2 $,
where $E_{\rm k}$ is the average kinetic energy for the center of mass motion.
Using the thermal RMS kinetic energy, this gives a predicted average splitting
of $\approx 150$\,MHz.  This naturally explains the shape of the
predicted spectrum and the energy scale.  

\begin{figure}
\centerline{\epsfbox{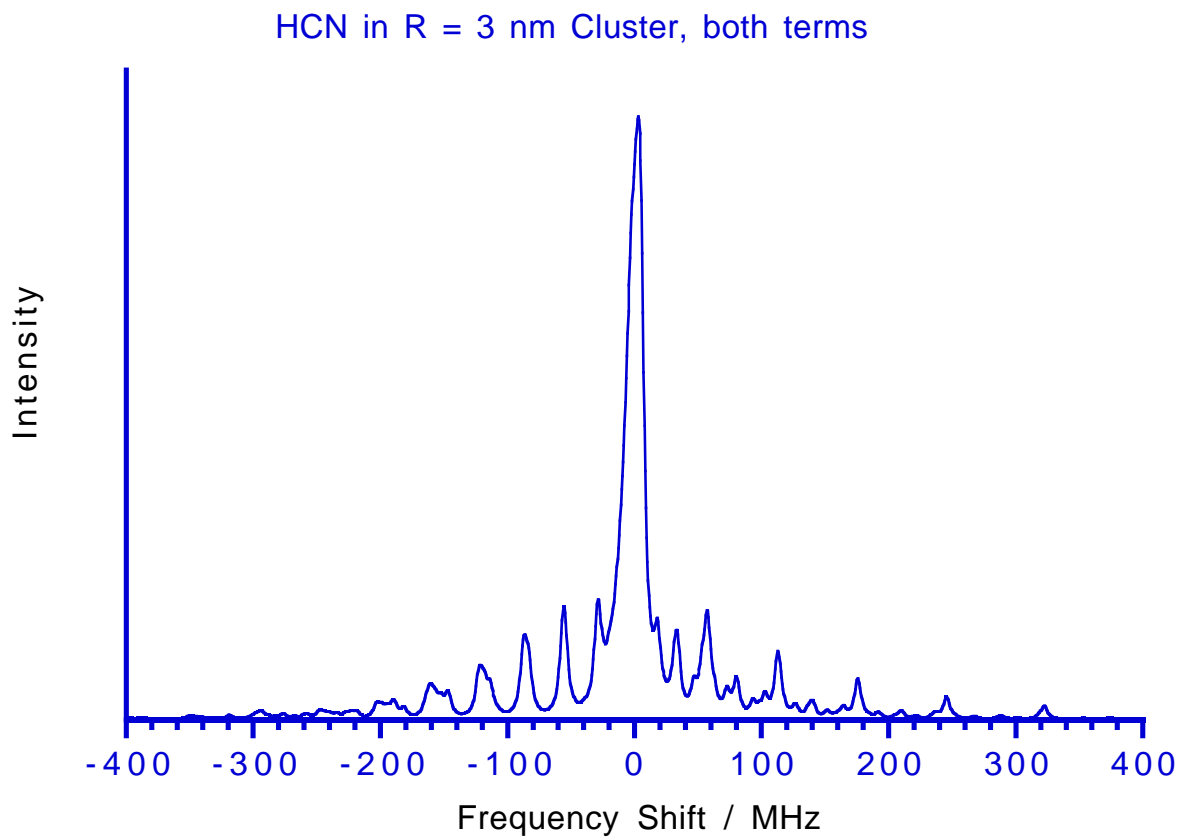}}
\vspace{1in}
\label{HCN_spec3}
\caption{Calculated HCN R(0) line, including the isotropic
and anisotropic terms in the potentials and the hydrodynamic coupling, 
for a He cluster of $R = 3$\,nm.}
\end{figure}

Figure~\ref{HCN_spec3} shows the calculated cluster spectrum for the HCN
R(0) line including both the anisotropic potential term and the
hydrodynamic coupling.  Somewhat surprisingly, the spectrum is narrower
than for the case of either coupling term by itself.  The reason is that
for the dominant $\Delta L = 0$ matrix elements, the two sets of matrix
elements destructively interfere.  This is despite the fact that the two
coupling terms are complementary in that the hydrodynamic coupling is most
important when the impurity is near the center of the cluster (and thus
has the highest velocity), while the potential anisotropy is most important
at the outer turning point of the motion.
The quantitative effect of this interference on the predicted spectrum will
be a strong function of the particular parameters used in the calculation, 
including the cluster size.

\section{Application to the R(0) transition of OCS}\label{sec-OCS}
 
We will now consider the application of the present model to
the case of OCS as an impurity, partly because of the 
elegant studies of the spectrum of this molecule performed in G{\"o}ttingen,
and also because it has the narrowest lines presently observed for
an impurity in $^4$He clusters~\cite{Hartmann_thesis}.
In fact, the linewidths estimated above for the R(0) transition of
HCN, while being well below that observed 
in the R(0) transition HCN in the first C--H overtone band,
are similar to that of the R(0) line of the OCS $\nu_3$
fundamental.  This suggests that the present model may be
more accurate in this case. 

At the author's request, Joanna Howson and Jeremy Hutson 
have calculated long range
coefficients for the He--OCS pair~\cite{Howson_up}.
They find $C_{60} = 14.2\,{\rm eV \AA}^6$, $C_{62} = 2.44\, {\rm eV \AA}^6$,
$C_{71} = 14.4\,{\rm eV \AA}^7$ and $C_{73} = 13.9\,{\rm eV \AA}^7$.
Calculations were done with a number of basis sets, but the
above results were done with a Dunning augmented--cc--pVTZ basis and
are reported to have $\approx 10\%$ uncertainty.  The calculations use
SCF polarizabilities and employed the CADPAC6.3 program package.

In the present paper, we will continue to use only the lowest order, $C_6$,
terms, though the type of scaling done above indicates that the $C_{71}$
term may make a substantial contribution in the present case.      
Higgins and Klemperer~\cite{Higgins99} have calculated
a series of points on the He--OCS surface at the MP4 level of theory,
which they kindly made available to the author.  
From the position of the minimum as a function of angle between the
He--OCS vector and the OCS molecular axis, we approximate the molecule as
an ellipsoid with $a = b = 3.4\,{\rm \AA}$ and $c = 4.7\,{\rm \AA}$, or $\xi = 1.38$.

\begin{figure}
\epsfxsize=5in
\centerline{\epsfbox{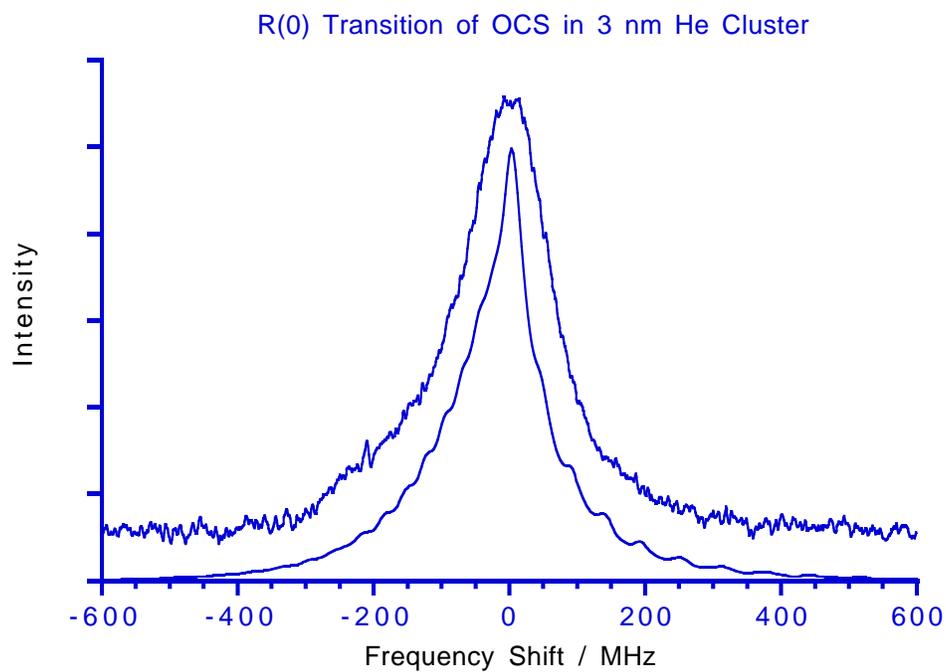}}
\vspace{1in}
\label{OCS_a}
\caption{Plot of R(0) line of the $\nu_3$ fundamental band of OCS
in a $^4$He cluster of mean size $N = 2700$ ($R = 3.1$\,nm).
Taken from the thesis of M. Hartmann.  Also included is the
calculated lineshape using anisotropic potential and 
hydrodynamic coupled, as decribed in the text, convoluted
with a Lorentzian of $30$\,MHz FWHM.}
\end{figure}

Figure~\ref{OCS_a} shows the R(0) transition of OCS observed in 
$^4$He clusters, as reported by Grebenev {et al.}~\cite{Grebenev99}.
It has a FWHM of $150$\, MHz, and is asymmetric with a broad
shoulder going to lower frequencies. Also in Figure~\ref{OCS_a} is plotted
the R(0) transition for OCS, predicted from the above parameters,
and convoluted with a Lorentzian of $30$\, MHz FWHM 
to wash out most of the fine structure in the predicted spectrum.
This is justified both because we expect vibrational and
rotational relaxation lifetime to broaden the lines, and because
the average over the cluster size distribution should at least partially
wash out the 
high resolution structure.  The agreement with experiment is 
remarkable, especially considering the uncertainty in the parameters
going into the model.  The calculated linewidth is somewhat too
small, but the shape of the peak is in excellent agreement with what
was observed in the experiment.
 
This level of agreement is highly suggestive that the interactions
included in the present model are the principal source of line
broadening in the OCS R(0) transition.  However, it must be noted that 
the experiments have shown that the linewidth of the ro--vibrational lines
increases dramatically with J, with the R(0) line being the narrowest.
While the author has not calculated lineshapes for higher J transitions
because of memory limitations on the personal computer on which these
calculations have been performed, it is likely that the present model will
not predict an increase in linewidth with J, but will more likely predict 
the opposite.  Also the experiments show a significant difference in linewidth
for the R(0) and P(1) lines, which should have the same lineshape (except for
an inversion) in the present model, and in fact any model that does not
include a vibrational dependence to the He--impurity interaction.  We 
conclude that while the interactions considered in this paper likely
play an important role in the lineshape of the R(0) transition, as yet
unidentified interactions also play an important role, especially at higher J.

\section{Hydrodynamic Contribution to the effective moment of Inertia}\label{sec-Inertia}

\begin{figure}
\centerline{\epsfbox{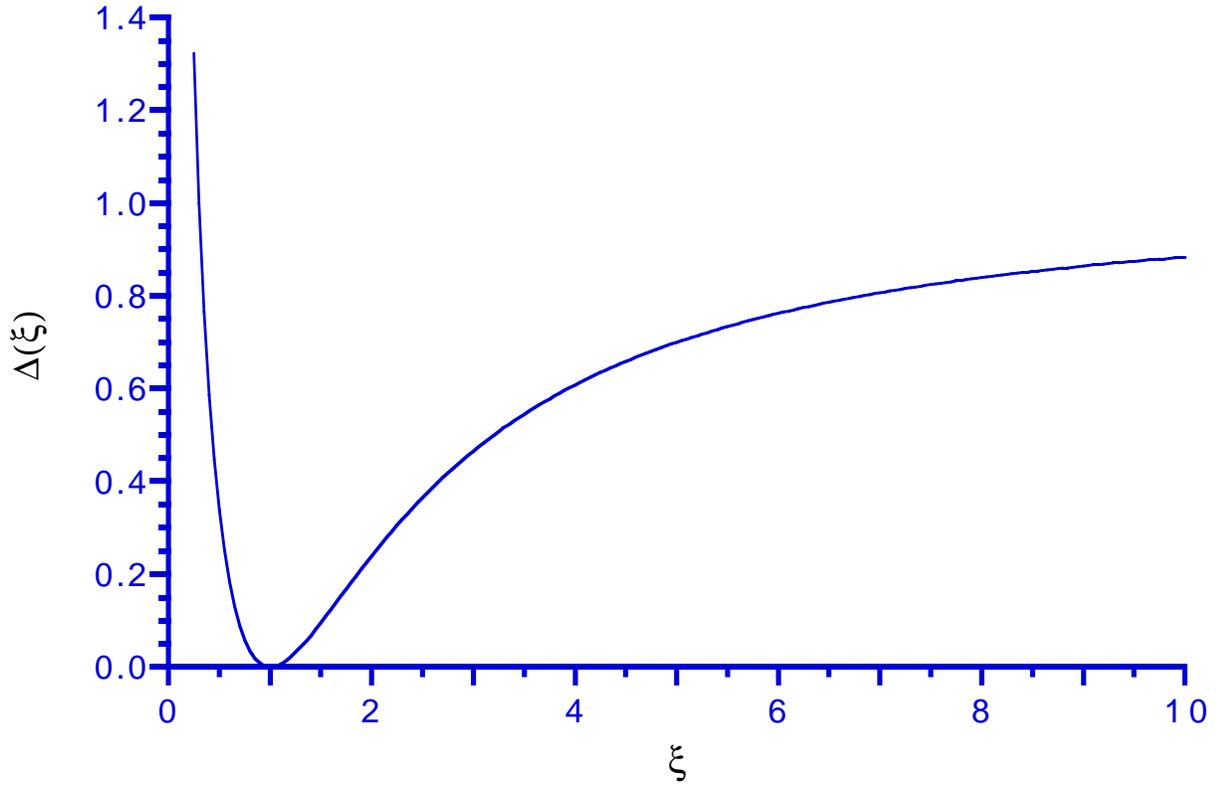}}
\vspace{1in}
\label{Hydro_2}
{\caption{The hydrodynamic constant  $\Delta(\xi)$ as a function of $\xi$.
See Eq.~\ref{eq:Delta} for the definition of this quantity.
}}
\end{figure}

Let us now consider the hydrodynamic effect on the effective moment
of inertia for rotation of our symmetric top impurity.  It has
been established experimentally that molecules rotate in 
liquid He clusters with
an effective rotational constant that is smaller than that of the
same molecule in the gas phase.  This implies an increase in the effective
moment of inertia for rotation.  For light impurities with large rotational
constants, fractional change of the rotational constants is small.  
For example, in the case of H$_2$O, the rotational constants in
the cluster are almost identical with the gas phase~\cite{Frochtenicht96}.  
In contrast, heavy impurities with small gas phase rotational 
constants, such as $\rm (CH_3)_3SiCCH$ have a rotational constant in the
cluster of only $\approx 20\%$ as large as in the gas 
phase~\cite{Callegari_up}.  
Some reduction in rotation constant is expected because of
hydrodynamic effects, i.e., He must move out of the way of the rotating
impurity, which contributes to the kinetic energy.  Using the model of
the impurity as an ellipsoid, we can directly calculate this increased
moment of inertia from the size and shape of the molecule. The 
velocity potential for such motion is given by Milne-Thomson.  Only
an integral over the surface of the ellipsoid is needed to calculate
the hydrodynamic contribution to the rotational kinetic energy.  This leads
to the following result for the hydrodynamic
contribution to the moment of inertia for rotation 
about the axis perpendicular to
the symmetry axis of the ellipsoid:
\begin{eqnarray}
\Delta I_{B} &=& I' \Delta(\xi) + M' \gamma_1 r_{com}^2 \\
I' &=& \frac{1}{5} M' a^2 ( 1 + \xi^2) \\
\Delta(\xi) &=& \frac{1}{1 + \xi^2} \cdot \frac{(1-\xi^2)^2 (\alpha_0 - \gamma_0)}
{2(1-\xi^2) - (1 + \xi^2) (\alpha_0 - \gamma_0)} \label{eq:Delta}
\end{eqnarray}
$r_{com}$ is the displacement of the impurity center of mass from
the geometric center of the ellipsoid and 
$I'$ is the value of $I_B$
for the ellipsoid if filled with liquid He and this provides 
the natural scale for the effect.  
The other symbols are as defined above.  
Figure~\ref{Hydro_2} shows a plot of $\Delta(\xi)$ as a function
of $\xi$.  For $\xi \approx 1$, one can use the expansion
$\Delta(\xi) \approx \frac{2}{3} (\xi - 1)^2 \dots$.

As an application of this expression, we will again consider OCS.
Using the parameters that $a = b = 3.4 {\rm \AA}$ and  
$c = 4.7 {\rm \AA}$ estimated from the 
{\it ab initio} points~\cite{Higgins99}, and
$r_{com} = 0.1 {\rm \AA}$, the calculated enhancement of the B
moment of inertia is $9.33\, {\rm u \AA}^2$, compared to 
a value of $I_B = 83.1\, {\rm u \AA}^2$, or a predicted
rotational constant of OCS in the cluster of $90 \%$ of the 
gas phase value. This is a small fraction of the observed
reduction in B, where the observed
B value in He clusters is $36\%$ of the 
gas phase B value~\cite{Hartmann_thesis}.

Another model for the reduction of rotational
constants is to assume a solvation shell of He atoms that rotate
with the impurity.  In the case of SF$_6$, such a model seems to
naturally explain the observed reduction in B~\cite{Hartmann96},
but does not naturally explain the observed value of the A
rotational constant for (SF$_6$)$_2$~\cite{Hartmann_thesis},
which by such a model would be expected to be at least $1/2$ of
the B value for the SF$_6$ monomer in the cluster, while
the observed value (based upon a spectral fit) is less than
$1/4$.  In their recent paper on OCS in mixed $^3$He and $^4$He clusters, 
Grebenev {\it et al.} ~\cite{Grebenev98} propose a model for
the effective rotational constants of impurities in He with
a position dependent normal fluid density around the impurity that
rotates with it and thus contributes to the moment of inertia.
However, they give no microscopic description of how this normal
fraction could be defined, nor how it could be measured, thus this
model appears to lack empirical content at least until it is
more precisely defined.  It is interesting to point out in
this context that previous fully
quantum microscopic calculations have found that He atoms in the 
first solvation
shell about SF$_6$ participate in exchange and are thus part of the 
superfluid~\cite{Kwon96}.  These authors did find that at 
0.625 K the superfluid fraction in a SF$_6$He$_{39}$, 0.67(7),
was less than previously
calculated for a He$_{64}$ cluster, 0.9(1). It would be
interesting for these calculations to be repeated for larger He clusters
(at least enough to `fill' the second solvation shell around the SF$_6$),
and at a temperature near that found for experimental clusters in order
to make better contact with the presently available spectroscopic experiments.

\section{Summary}\label{sec-Summary}

In this paper, we have considered the energetics and dynamics for the
motion of impurities in spherical liquid $^4$He clusters, which are treated
as superfluid liquid droplets.  It is found that the effective potential
for motion of the impurity is determined by the long range part of the 
impurity--He interaction.  The isotropic part `traps' the impurity near
the center of the cluster.  The anisotropic part results in a coupling
of the rotation and center of mass motion of the impurity, which leads to 
inhomogeneous broadening in the predicted IR and microwave spectra.  
Calculated broadening values are on the order of 0.1--1 GHz, on the order of the
narrowest lines observed in IR spectroscopy of impurities in liquid He.  
Hydrodynamic effects are also found to play an important role by coupling
molecular rotation to the velocity of center of mass motion. 
The inhomogeneous spectral effects of the distribution of thermally excited
ripplons are found to be much smaller and are likely negligible in 
determining linewidths, though they likely play an important role in 
determining the time required for the impurity and cluster to come
into equilibrium.  Direct hydrodynamic effects are found to make a minor
contribution to the dramatic reductions in effective rotational constants
often found for impurity species in He cluster spectroscopy. 

This work will surely not be the last word on this subject.  The
interactions considered here, while certainly present, do not appear to
be sufficiently strong to explain the observed linewidths of many impurities, 
such as SF$_6$ and HCN.  This suggest other, perhaps more subtle, 
physical effects are playing a role.  New nonlinear spectroscopic experiments
that use either pump--probe methods and/or spectral hole burning are clearly
needed to sort out homogeneous and inhomogeneous effects in He cluster
spectroscopy and to provide critical tests of simple models.  
Microscopic theory will also surely play an important role in helping to
sort out the various physical effects that can play a role.  

\section*{Acknowledgments}

The He cluster research at Princeton is carried with support from the
from the Air Force High Density Materials program and the
National Science Foundation.
The author would like to acknowledge Prof. Giacinto Scoles for may helpful 
discussions about intermolecular potentials and for inviting the author to
join his research effort on He clusters.  
Kelly Higgins and William Klemperer as well as
J.~P.~Toennies and his co--workers are acknowledged for making
their results available prior to publication.  Joanna Howson and
Jerry Hutson are noted for kindly calculating the long range interaction
constants for He--OCS.  
Carlo Callegari, Irene Reinhard and Andrej Vilesov are noted for their careful reading of
this manuscript, which helped greatly to improve the final version.
The hospitality and support of JILA, where the
work was completed and the paper written, is also acknowledged.

\newpage
\begin{table}[t] 
\caption{Thermodynamic averages
for the motion of SF$_6$ and ripplons in a He cluster at $T_{\rm c} = 0.38$\,K.
$R$ is the radius of the cluster, $Q$ the partition function for the center of 
mass motion
or the ripplons, $<E>$ is the thermally averaged energy, expressed in units of 
$k_b T_{\rm c}$, $C_v$ is the heat capacity in units
of $k_b$, $S$ is the entropy in units of $k_b$, $<${\boldmath $L$}$^2>$ is the mean squared
value for the thermal angular momentum, $<a^2>$ is the mean squared value for the
displacement of the impurity from the center of the cluster,
and $<n>$ is the mean number of quanta of excitation of the ripplon modes. }

\vspace{0.5in}
\begin{tabular}{|c|c|c|c|c|c|c|c|c|} \hline
  & $R$/nm & $Q$ & $\langle E \rangle /(k_b T_{\rm c})$ 
& $C_v / k_b $ & $S/k_b$ & $\sqrt{ \langle \mbox{\boldmath $L$}^2 \rangle }$ &
$\sqrt{ \langle a^2 \rangle}/$nm & $\langle n \rangle$ \\ \hline 
SF$_6$ Quantum   & 3.0   & 478. & 2.49 & 2.01 & 8.66 & 16.3 & &\\ \hline 
SF$_6$ Classical & 3.0   & 536. & 2.44 & 2.13 & 8.72 & 16.4 & 0.96 &\\ \hline 
Ripplons  & 3.0 & $8.7\cdot10^3$& 18.6 & 45.4 & 30.0 & 15.2 & & 14.8
\\ \hline \hline
SF$_6$ Classical & 5.0 & $8.2\cdot10^4$& 2.21 & 2.06 & 11.2 & 39.2 & 2.9 &
\\ \hline 
Ripplons & 5.0   & $1.1\cdot10^{16}$   & 54.1 &127.8 & 91.1 & 41.0 & & 54.9
\\ \hline
\end{tabular}
\end{table}


\end{document}